\documentclass[sigconf,screen,authorversion,pbalance]{acmart}

\usepackage{siunitx}
\usepackage{graphicx}

\usepackage{xspace}

\usepackage[nameinlink,noabbrev,capitalise]{cleveref}

\newcommand{\CLITOOL}[1]{\texttt{#1}}

\newcommand{\huffcode}{Huffman code\xspace}
\newcommand{\huffcodes}{Huffman codes\xspace}
\newcommand{\dynblocks}{Dynamic Blocks\xspace}
\newcommand{\dynblock}{Dynamic Block\xspace}
\newcommand{\fixedblocks}{Fixed Blocks\xspace}
\newcommand{\fixedblock}{Fixed Block\xspace}
\newcommand{\rawblocks}{Non-Com\-pressed Blocks\xspace}
\newcommand{\rawblock}{Non-Com\-pressed Block\xspace}
\newcommand{\deflateblock}{Deflate block\xspace}
\newcommand{\deflateblocks}{Deflate blocks\xspace}
\newcommand{\backrefwindow}{window\xspace}
\newcommand{\backrefwindows}{windows\xspace}
\newcommand{\blockfinder}{block finder\xspace}
\newcommand{\chunkfetcher}{\texttt{Chunk\-Fetcher}\xspace}
\newcommand{\tar}{TAR\xspace}
\newcommand{\gzip}{\CLITOOL{gzip}\xspace}
\newcommand{\pugz}{\CLITOOL{pugz}\xspace}
\newcommand{\pigz}{\CLITOOL{pigz}\xspace}
\newcommand{\bgzip}{\CLITOOL{bgzip}\xspace}
\newcommand{\pzstd}{\CLITOOL{pzstd}\xspace}

\newcommand{\pragzipname}{our implementation\xspace}
\newcommand{\deflate}{Deflate\xspace}
\newcommand{\precode}{Precode\xspace}

\newcommand{\pragzip}{\CLITOOL{rapidgzip}\xspace}
\newcommand{\Pragzip}{Rapidgzip\xspace}

\copyrightyear{2023}
\acmYear{2023}
\setcopyright{acmlicensed}\acmConference[HPDC '23]{Proceedings of the 32nd International Symposium on High-Performance Parallel and Distributed Computing}{June 16--23, 2023}{Orlando, FL, USA}
\acmBooktitle{Proceedings of the 32nd International Symposium on High-Performance Parallel and Distributed Computing (HPDC '23), June 16--23, 2023, Orlando, FL, USA}
\acmPrice{15.00}
\acmDOI{10.1145/3588195.3592992}
\acmISBN{979-8-4007-0155-9/23/06}

\begin{document}

\title{\Pragzip: Parallel Decompression and Seeking in Gzip Files Using Cache Prefetching}

\acmSubmissionID{\#65}

\author{Maximilian Knespel}
\email{maximilian.knespel@tu-dresden.de}
\orcid{0000-0001-9568-3075}
\affiliation{%
  \institution{Technische Universität Dresden}
  \department{Center for Information Services and High Performance Computing}
  \city{Dresden}
  \state{Saxony}
  \country{Germany}
}

\author{Holger Brunst}
\email{holger.brunst@tu-dresden.de}
\orcid{0000-0003-2224-0630}
\affiliation{%
  \institution{Technische Universität Dresden}
  \department{Center for Information Services and High Performance Computing}
  \city{Dresden}
  \state{Saxony}
  \country{Germany}
}

\begin{abstract}
Gzip is a file compression format, which is ubiquitously used.
Although a multitude of gzip implementations exist, only pugz can fully utilize current multi-core processor architectures for decompression.
Yet, pugz cannot decompress arbitrary gzip files.
It requires the decompressed stream to only contain byte values 9--126.
In this work, we present a generalization of the parallelization scheme used by pugz that can be reliably applied to arbitrary gzip-compressed data without compromising performance.
We show that the requirements on the file contents posed by pugz can be dropped by implementing an architecture based on a cache and a parallelized prefetcher.
This architecture can safely handle faulty decompression results, which can appear when threads start decompressing in the middle of a gzip file by using trial and error.
Using 128 cores, our implementation reaches \SI{8.7}{\giga\byte/\second} decompression bandwidth for gzip-compressed base64-encoded data, a speedup of 55 over the single-threaded GNU gzip, and \SI{5.6}{\giga\byte/\second} for the Silesia corpus, a speedup of \num{33} over GNU gzip.
\end{abstract}

\ccsdesc[500]{Theory of computation~Shared memory algorithms}
\ccsdesc[300]{Theory of computation~Data compression}
\ccsdesc[500]{Software and its engineering~Software performance}

\keywords{Gzip; Decompression; Parallel Algorithm; Performance; Random Access}

\maketitle

\section{Introduction}

Gzip~\cite{RFC1952} and \deflate~\cite{RFC1951} are compression file formats that are ubiquitously used to compress single files and file agglomerates like \tar archives, databases, and network transmissions.
A gzip file contains one or more gzip streams each containing a Deflate stream and additional metadata.
Each \deflate stream contains one or more \deflateblocks.
Three kinds of \deflateblocks exist: \rawblocks, \dynblocks, and \fixedblocks.
\dynblocks, and \fixedblocks are compressed using a combination of Huffman coding~\cite{huffman} and a variation of the Lempel–Ziv–Storer–Szymanski (LZSS) algorithm~\cite{lzss}.
LZSS is a variation of LZ77~\cite{lz77}, which compresses substrings with backward pointers to the preceding stream.

Deflate streams can also be found in other file formats like ZIP~\cite{zip}, PNG~\cite{png} and other ZIP-based file formats like JAR~\cite{jar}, XLSX~\cite{xlsx} and ODT~\cite{odt}.

\subsection{Motivation}

Gzip-compressed TAR files can grow to several terabytes, e.g., the updated \SI{1.19}{\tera\byte} large gzip-compressed ImageNet archive~\cite{imagenet21k}.
The multiple petabytes large Common Crawl~\cite{commoncrawl} dataset is also distributed as a set of gzip-compressed files.
A pipeline for decompressing and preprocessing such data, e.g., for use in machine learning context, would benefit from being accelerated by multi-threaded decompression.
Extracting a \SI{1}{\tera\byte} large file with \gzip takes several hours.
Assuming a sufficiently fast file system and two 64-core processors, this can be reduced by a factor of \num{55} to several minutes when using \pragzipname.

Gzip files are only slowly getting replaced by newer compression formats like Zstandard~\cite{zstandard} because of the wide-spread availability of ZLIB~\cite{zlib} and \gzip for compressing and decompressing gzip streams.
Parallel decompression of gzip becomes desirable when working with gzip-compressed data from external sources because the compression format cannot be chosen.
If the compression format can be chosen and if decompression support is not an issue, then newer formats like Zstandard are more efficient.
However, as we show in \cref{sct:compression-formats}, decompression with \pragzipname using 128 cores can be twice as fast as the parallel Zstandard decompression tool \pzstd given the same amount of cores.

\subsection{Limitation of State-of-Art Approaches}
\label{sct:limitations}

Although a multitude of implementations for gzip exist, none of them can fully utilize current multi-core processor architectures for decompressing gzip-compressed files without restrictions on the contents.
The Blocked GNU Zip Format (BGZF)~\cite{htslib} is a subset of gzip files that adds the encoded size of the gzip stream into the metadata of the gzip stream header.
With this metadata, another thread can seek to the next gzip stream and start decoding it in parallel.
The command line tool \CLITOOL{bgzip} uses multi-threading to decompress BGZF files.
It cannot parallelize the decompression of gzip files that do not contain the metadata defined by BGZF.

\citeauthor{pugz}~\cite{pugz} show that gzip decompression can be split into two stages, each of which can be parallelized.
For parallelization, the input data is divided into chunks, which are sent to decompression threads.
In the first stage, each thread searches for the first \deflateblock in its assigned chunk and starts first-stage decompression from there.
Searching for \deflateblocks may result in false positives, i.e., positions in the stream that are indistinguishable from a valid \deflateblock without parsing the whole preceding compressed stream.
The output of the first-stage decompression may contain markers for data referenced by backward pointers that cannot be resolved without knowing the preceding decompressed data.
These markers are resolved in the second stage.

There are some limitations to this approach:
\begin{itemize}
    \item Parallelization works on a \deflateblock granularity.
          Gzip streams containing fewer \deflateblocks than the degree of parallelization aimed for cannot be effectively parallelized.
    \item False positives for \deflateblocks cannot be prevented without knowing the preceding \deflate stream.
    \item There is overhead for the two stages, the intermediate format, which is twice the size of the result, and for finding new blocks.
          The larger the overhead for finding a block is, the larger the chunk size has to be chosen to achieve a speedup close to ideal linear scaling.
          This imposes a lower bound on the chunk size for reaching optimal computational performance.
    \item The memory usage is proportional to the degree of parallelism, the chunk size, and the compression ratio because the decompressed data of each chunk has to be stored in memory.
\end{itemize}

An implementation of this algorithm exists in the command line tool \pugz.
To reduce false positives when searching for \deflateblocks, it requires that the decompressed data stream only contains bytes with values 9--126 and decompresses to at least \SI{1}{\kibi\byte} and up to \SI{4}{\mebi\byte}.
It is not able to decompress arbitrary gzip files.

Furthermore, chunks are distributed to the parallel threads in a fixed uniform manner and therefore can lead to workload imbalances caused by varying compression factors in each chunk.
Because it is based on libdeflate~\cite{libdeflate}, it also has the technical limitation that the output buffer size for each chunk has to be configured before decompression.
The output buffer is set to \SI{512}{\mebi\byte} of decompressed data for each of the \SI{32}{\mebi\byte} chunks in the compressed data stream, meaning it will fail for files with a compression ratio larger than \num{16}.

\subsection{Key Insights and Contributions}

Our implementation, \pragzip\anon[\footnote{Name has been changed for double blind review}]{}, addresses the limitations of \pugz by implementing a cache and a parallelized prefetcher.
It is open-source software and available on Github\footnote{\anon[URL removed for double-blind review]{\url{https://github.com/mxmlnkn/indexed_bzip2}}}.
Our architecture solves the following issues:
\begin{itemize}
    \item False positives that are found when searching for \deflate blocks are inserted into the cache but are never used and will be evicted.
          This makes the architecture robust against false positives and enables us to generalize gzip decompression to arbitrary gzip files by removing the non-generalizing \deflate block finder checks.
    \item The prefetched chunks are pushed into a thread pool for workload balanced parallel decompression.
    \item Non-sequential access patterns are supported efficiently.
    \item Gzip files with more than one gzip stream are supported.
    \item Abstraction of file reading enables support for Python file-like objects.
\end{itemize}

Caching and prefetching are techniques commonly used in processors to optimize main memory access~\cite{smith1982cache}, file access~\cite{mpi-io-caching}, and other domains.
We have successfully applied these techniques for chunks of decompressed data.
This enables us to provide not only parallelized decompression but also constant-time random access into the decompressed stream given an index containing seek points.
While a proof-of-concept for parallel decompression for specialized gzip-compressed data was available with \pugz~\cite{pugz} and random access into decompressed gzip streams was available with \texttt{indexed\_gzip}~\cite{indexed_gzip}, we have successfully combined these two capabilities and made them usable for arbitrary gzip files.

\paragraph{Index for Seeking}
An index containing seek points is built during decompression.
Each seek point contains the compressed bit offset, the decompressed byte offset, and the last \SI{32}{\kibi\byte} of decompressed data from the previous block.
Decompression can be resumed at each seek point without decompressing anything before it.
The range between the previous seek point and the requested offset is decompressed in order to reach offsets between seek points.
The overhead for this is bounded because the seek point spacing can be configured to a maximum value.

Our implementation's seeking and decompression capabilities can also be used via a library interface to provide a light-weight layer to access the compressed file contents as done by ratar\-mount \cite{ratarmount}.
The seek point index can be exported and imported similarly to \texttt{indexed\_gzip}~\cite{indexed_gzip} to avoid the decompression time for the initial decompression pass.
Furthermore, decompression can be delegated to an optimized gzip implementation like zlib~\cite{zlib} when the index has been loaded.
This is more than twice as fast as the two-stage decompression.
Lastly, loading the index also improves load balancing and reduces memory usage because the seek points are equally spaced in the decompressed stream.

\paragraph{Performance Analysis}

We also provide benchmark results for all components of \pragzipname in \cref{sct:evaluation} in order to give a performance overview and determine bottlenecks.
We improve one such bottleneck, the searching for \deflate blocks, by implementing a skip table.
This block finder is 4 times faster than the one implemented in \pugz.
A faster block finder makes it possible to reduce the chunk size and, therefore, the memory requirements accordingly while retaining the same overall performance.
The fast \deflate \blockfinder also improves the speed for the recovery of corrupted gzip files.

\subsection{Limitations of the Proposed Approach}

The main limiting factor of \pragzipname is memory usage.
Each chunk of decompressed data currently being prefetched and contained in the cache is held in memory.
The prefetch cache holds twice as many chunks as the degree of parallelization.
The compressed chunk size can be configured and is \SI{4}{\mebi\byte} by default.
During index creation, large chunks are split when necessary to ensure that the maximum decompressed chunk size is not larger than the configured chunk size.
Therefore, the maximum memory requirement is only \SI{8}{\mebi\byte} per thread if an existing index is used for decompression.
The memory required for each chunk depends on the compression ratio of the file when decompressing without an existing index.
Compression factors are often below \num{10}, which translates to a memory requirement of \SI{80}{\mebi\byte} per thread.
In the worst case, for a file with the largest possible compression factor of \num{1032}, the memory requirement for decompression would be \SI{8.3}{\gibi\byte} per thread.
This can be mitigated by implementing a fallback to sequential decompression for chunks with large compression ratios or by splitting the chunk sizes dynamically.

A second limitation is that the achievable parallelization is limited by the number of \rawblocks and \dynblocks in the file.
If the file consists of a single gzip stream with a single \deflateblock, then \pragzipname cannot parallelize decompression.
This is a limitation of the two-stage decompression scheme.
In contrast to \pugz, the block finder in \pragzipname does not look for \deflate blocks with Fixed Huffman Codes.
If a file consists only of such \deflate blocks, then the decompression will not be parallelized with our implementation.
With default compression settings, such blocks occur only rarely, namely for very small files and at the end of a gzip stream.
In \cref{sct:gzip-compressors}, we show that most gzip compression tools and compression level settings result in gzip files that can be decompressed in parallel with \pragzipname.

\section{Theoretical Background}
\label{sct:theory}
In this section, we summarize the \deflateblock format~\cite{RFC1951} to explain the issue of backward pointers hindering parallelization and to explain our block finder optimizations.
We also give a short description of the two-stage \deflate decompression.

\subsection{The Deflate Stream Format}

\begin{figure}
    \centering
    \includegraphics[width=0.5\linewidth]{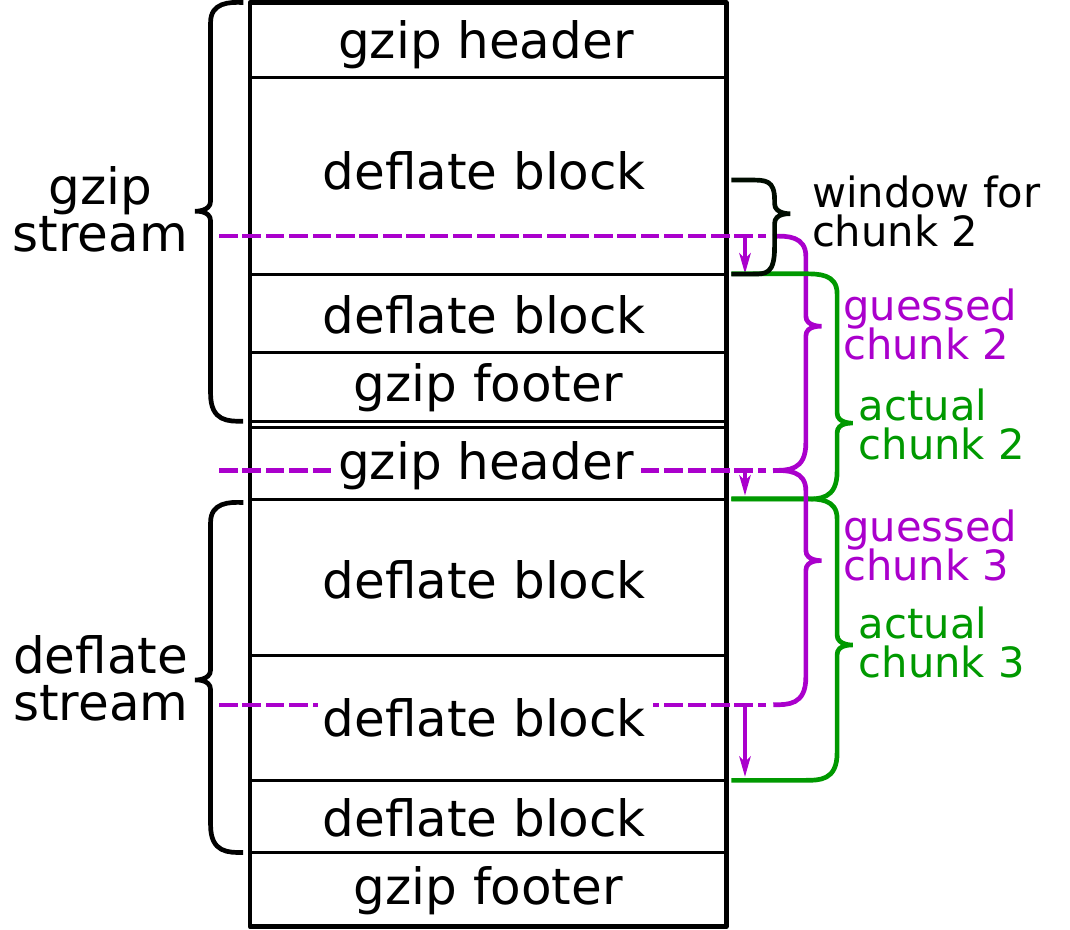}
    \caption{
        The structure of a gzip file. Also marked are the chunks as they are assigned to the decompression threads and the first \deflateblock in each chunk, i.e, the actual chunk start as it will be written to the index.
    }
    \label{fig:gzip-file}
\end{figure}

Even though the presented parallel decompressor works on gzip files, most implementation details are focused on the \deflate format.
The gzip file format~\cite{RFC1952}, shown in \cref{fig:gzip-file}, consists of one or more gzip streams.
A gzip stream wraps a raw \deflate stream and adds metadata such as file format identification bytes, the original file name, and a checksum.

\begin{figure}
    \centering
    \begin{minipage}{0.49\linewidth}\begin{center}
        \textbf{\rawblock}\\[8pt]
        \includegraphics[scale=1]{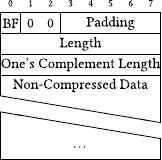}\\[8pt]
        \textbf{\fixedblock}\\[10pt]
        \includegraphics[scale=1]{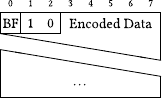}
    \end{center}\end{minipage}\begin{minipage}{0.49\linewidth}\begin{center}
        \textbf{\dynblock}\\[8pt]
        \includegraphics[scale=1]{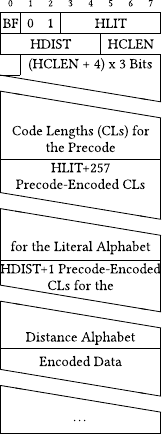}
    \end{center}\end{minipage}
    \caption{
        Deflate block formats.
        A \deflateblock can start at any bit offset.
        \rawblocks add padding bits after the block type to byte-align the length fields and the subsequent data.
        The block is the last one in the \deflate stream if the final-block (BF) bit is set.
        The block type value $\mathbf{{11}_2}$ is reserved.
    }
    \label{fig:deflate_blocks}
\end{figure}

\Cref{fig:deflate_blocks} shows the three types of \deflate blocks:
\begin{itemize}
    \item Non-Compressed: Contains a stream of original data. Used for incompressible data.
    \item Compression with fixed Huffman codes: Contains data compressed with a predefined \huffcode.
          Used for small data to save space by not storing a custom \huffcode.
    \item Compression with dynamic Huffman codes: Contains a \huffcode in its header followed by data compressed with it.
\end{itemize}
These blocks will be referred to in shortened forms as \rawblock, \fixedblock, and \dynblock.
\deflate blocks are concatenated to a \deflate stream without any byte-alignment.

Each \dynblock uses three different Huffman codes (HC): one is for the Literal alphabet, the second one for the Distance alphabet, and the third one is to encode the code lengths (CL) for defining the first two Huffman codes themselves also referred to as Precode.
The size of each HC is stored in HLIT, HDIST, and HCLEN respectively as shown in \cref{fig:deflate_blocks}.

The steps for decompressing a \dynblock are as follows:
\begin{enumerate}
    \item Read the lengths HLIT, HDIST, and HCLEN.
    \item Read the code lengths for the Precode and build a Huffman code from them.
    \item Use the \huffcode defined by the Precode to read the code lengths for the Distance alphabet and the Literal alphabet.
    \item Build a Huffman code from the Distance alphabet code lengths.
    \item Build a Huffman code from the Literal alphabet code lengths.
    \item Decode the \deflate data using the Literal and Distance alphabets.
\end{enumerate}

The compressed data encoded with the Literal alphabet contains either raw literals or pointers to duplicated strings, i.e., instructions to copy a given length of data from a given distance.
The distance is limited to \SI{32}{\kibi\byte} of decompressed data, i.e., backward pointers are limited to a sliding window.
Small lengths are encoded in the instructions and large lengths require reading more bits from the stream.
A distance encoded using the Distance \huffcode follows all backward pointers.

The instructions to copy sequences from the \backrefwindow introduce data dependencies that complicate parallelization.

\subsection{Two-Stage Deflate Decoding}
\label{sct:theory:two-stage-decompression}

\begin{figure}
    \centering
    \includegraphics[width=0.95\linewidth]{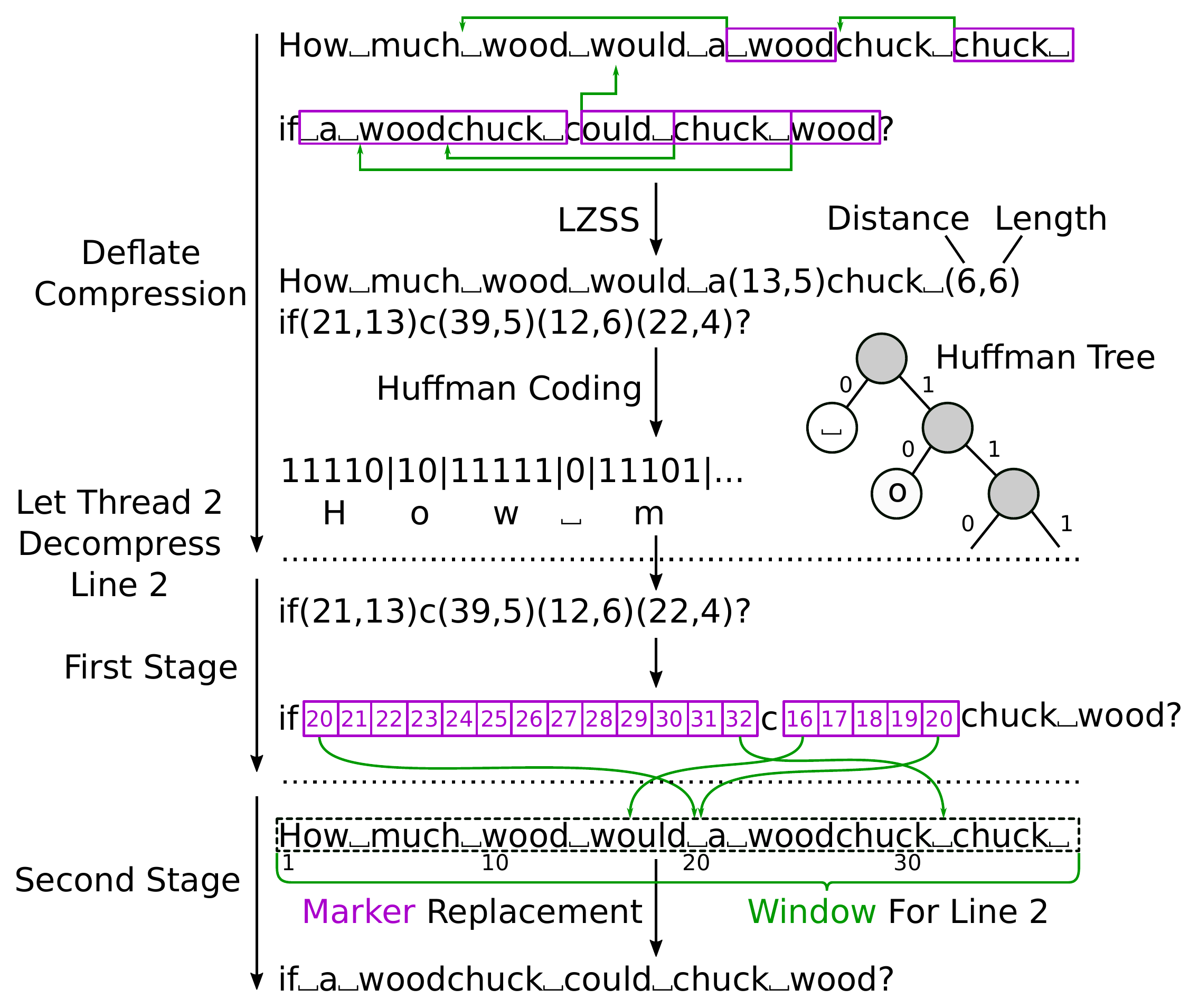}
    \caption{Example of \deflate compression and two-stage decompression.}
    \label{fig:two-stage-example}
\end{figure}

The authors of \pugz~\cite{pugz} show that it is feasible to parallelize decompression even if there are instructions to copy from an unknown \backrefwindow.
This is done by resolving placeholders for unknown values in a second step after those values have become known.
An example of this two-stage decompression is shown in~\cref{fig:two-stage-example}.
The steps for a decompression thread starting at an arbitrary offset in the file are:
\begin{enumerate}
    \item Find the next \deflateblock.
          Searching for \deflate blocks is a technique that has previously been used for reconstructing corrupted gzip files~\cite{park2008data}.
          This can be implemented by trying to decompress at the given offset and going to the next offset on error.
    \item Start decoding by filling the as-of-yet unknown \backrefwindow with unique 15-bit wide markers corresponding to the offset in the buffer.
          The decompression routine has to be adjusted to output an intermediate format with 16-bit symbols to store the markers itself or a literal and an additional bit to signal whether it is a marker or a literal.
    \item After the \backrefwindow has become available, replace all marker symbols with data from the \backrefwindow to get the fully decompressed contents.
\end{enumerate}

The benchmark results in \cref{sct:evaluation} show that the marker replacement is a magnitude faster than \deflate decompression.
This means that even if only the first stage has been parallelized and the marker replacement is applied serially, then decompression would be up to a magnitude faster. %
The \backrefwindows can be propagated even faster because the marker replacement only has to be applied for the last \SI{32}{\kibi\byte} of each chunk.
The remaining marker symbols can be replaced in parallel such that each chunk is processed by a different thread.
The propagation of the windows cannot be parallelized.
Assuming a chunk size of \SI{4}{\mebi\byte}, propagating only the last \SI{32}{\kibi\byte} of each chunk would be \num{128} times faster than replacing the markers in the whole chunk.
However, the non-parallelizable part takes up half the time when using \num{128} cores.
Therefore, the maximum achievable speedup depends on the chunk size and has an upper bound according to Amdahl's law.

\section{Implementation}
\label{sct:implementation}

\begin{figure}
    \centering
    \includegraphics[width=\linewidth]{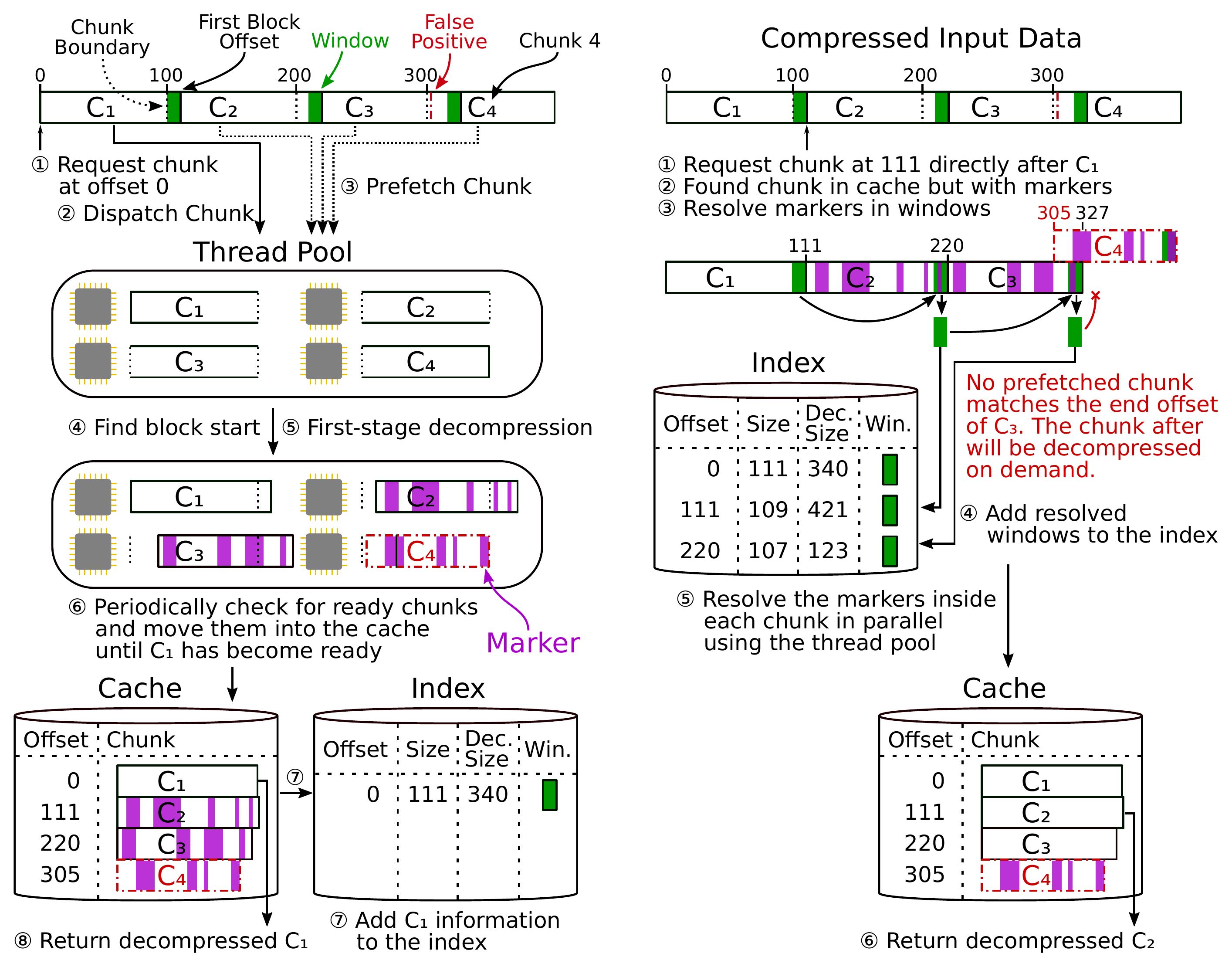}
    \caption{Example of parallel decompression with the cache-and-prefetch architecture.
             Accessing the first chunk will prefetch further chunks in parallel.
             On access to the next chunk, the cached result can be used after the markers have been replaced.}
    \label{fig:parallel-decompression-process}
\end{figure}

\begin{figure}
    \centering
    \includegraphics[width=\linewidth]{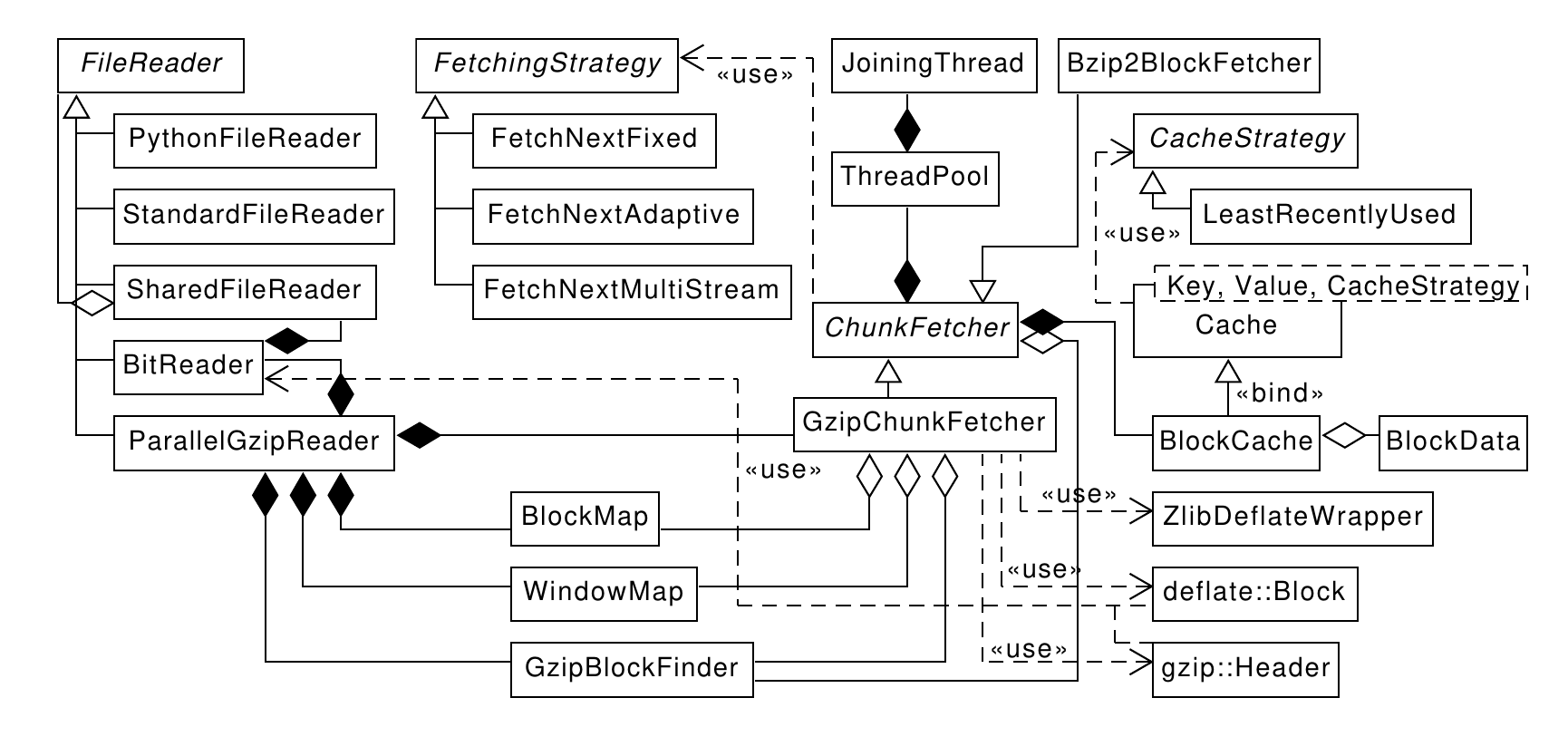}
    \caption{Class diagram of the main components of our implementation.}
    \label{fig:architecture}
\end{figure}

\Cref{fig:architecture} shows the class hierarchy of \pragzip. The main components are:
\begin{itemize}
    \item A chunk fetcher that can opaquely prefetch and cache chunks in parallel
    \item A \blockfinder that returns likely \deflateblock offsets in a given data stream
    \item A gzip/deflate decoder that supports two-stage decoding
    \item A database containing \deflateblock offsets and sliding windows to start decoding from
\end{itemize}

The \texttt{FileReader} interface abstracts file access to support not only file access to regular files but also access to Python file-like objects.
A Python program can use this feature to implement recursive access to the contents of gzip-compressed gzip files. %
Our implementation fulfills the following design goals:
\begin{itemize}
    \item parallel chunk decompression
    \item only requires an initial decompression pass until the requested offset
    \item fast concurrent access at two different offsets
    \item seeking is possible in constant time when the requested offset exists in the index
    \item building the index is not a preprocessing step, it is done on the fly
    \item robust against false positives returned by the block finder
\end{itemize}
Concurrent access at different offsets is common when providing access to gzip-compressed (TAR) files via a user-space filesystem, e.g., provided by \CLITOOL{ratarmount}~\cite{ratarmount}.
Therefore, our implementation not only generalizes the approach from \pugz~\cite{pugz} to non-ASCII contents but also incorporates random access capabilities similar to \texttt{indexed\_gzip}~\cite{indexed_gzip}.

Robustness against false positives results from the cache acting as an intermediary with the offset as key.
When a thread finds a false positive, then the wrong result will be inserted into the cache with a wrong offset as key.
The main thread will request the next chunk based on the end offset of the previous chunk.
If the prefetching of the next chunk found a false positive, then the main thread will not be able to find any matching chunk in the cache and has to dispatch a new decompression task for the previous chunk's end offset.

\subsection{Seeking and Reading}

A seek only updates the internal position and the end-of-file flag.
Any further work for seeking will be done on the next read call.

\Cref{fig:parallel-decompression-process} shows an example of parallel decompression when reading consecutively from the start of the file.
During a read call to the \texttt{ParallelGzipReader} class, the internal \texttt{GzipChunkFetcher} is requested to return the chunk that contains the given position.
Access to a chunk triggers the prefetcher, even if the chunk has been found in the cache.
If the chunk has not been cached, then a decompression task is dispatched to the thread pool on demand.
Cached chunks can contain marker symbols.
Marker replacement is also dispatched to the thread pool.
At this point, all necessary information for a seek point in the index is available and will be inserted.
A fully decompressed chunk is returned to the caller after the marker replacement task has finished.

\subsection{Prefetching}

Prefetching is applied according to a prefetch strategy and is computed on the chunk indexes, not the chunk offsets.
The \chunkfetcher class combines a thread pool, a cache, a prefetch cache, a prefetching strategy, and a database for converting chunk offsets to and from chunk indexes.
The prefetch cache is separate from the cache of accessed chunks to avoid cache pollution of the latter cache caused by the prefetching.
For the common case of full file decompression, the access cache size is set to one and will only contain the last accessed chunk.
\Cref{fig:parallel-decompression-process} displays the two caches as one for brevity.

The current default prefetching strategy is an ad-hoc algorithm that works for concurrent sequential access to one or more files in a gzip-compressed TAR archive.
It is comparable to an exponentially incremented adaptive asynchronous multi-stream prefetcher~\cite{gill2007amp}.
The prefetch strategy returns the full degree of prefetch for the initial access so that decompression starts fully parallel.
The prefetch strategy does not keep track of previously prefetched chunks.
It returns a list of chunk indexes to prefetch based on the last accessed chunk indexes.
The prefetcher has to filter out already cached chunks and chunks that are currently being prefetched.

\subsection{Chunk Decompression}

The \texttt{GzipChunkFetcher} class extends the \chunkfetcher class with the seek point index.
The index contains compressed offsets, uncompressed offsets, and the \backrefwindow for each chunk.
The \texttt{Gzip\-Chunk\-Fetcher} also provides the implementation for the chunk decompression task that is dispatched to the thread pool.
If a \backrefwindow exists in the index for a chunk offset, then the decompression task will delegate decompression to zlib.
If no \backrefwindow exists, then the custom-written \pragzip gzip decompressor will try to decompress at candidate offsets returned by the \blockfinder until decompression succeeds.
The decompressor will stop when a \dynblock or \rawblock that starts at or after the given stop offset has been encountered.
This stop condition excludes \fixedblocks because the \blockfinder does not look for those.
If the stop condition does not match the \blockfinder search conditions, then performance will degrade because the prefetched subsequent chunk offset will not be found in the cache and therefore has to be decompressed again.

\Pragzip implements full gzip parsing consisting of: gzip header parsing, deflate block parsing, and gzip footer parsing.
For \deflate parsing, it supplies the \texttt{deflate::Block} class, which supports two-stage decompression as well as conventional decompression when a \backrefwindow is given.
For \rawblocks, it contains a fast path that simply copies the raw data into the result buffer.
The \deflate decompressor also keeps track of the last non-resolved marker symbol.
Decompression will fall back to the faster conventional decompression if there is no marker symbol in the \backrefwindow.
This improves performance when a file contains large \rawblocks and for data compressed with only few backward pointers.

\subsection{Block Finder}

The \blockfinder returns the next \deflateblock candidate offset when given an offset from which to start searching.
It may return false positives and should but is not required to find all valid \deflate blocks.

False positives cannot be avoided when searching from an arbitrary offset.
When compressing a non-compressible gzip-file with gzip, the resulting gzip stream will consist of \rawblocks.
The verbatim parts of those blocks contain valid \deflate blocks of the gzip file that has been compressed.
These can be detected as false positives during parallel decompression.

The \blockfinder is split into specialized \deflateblock finders for \dynblocks and \rawblocks.
These two are combined by finding candidates for both and subsequently returning the result with the lower offset.

\subsubsection{Non-Compressed Blocks}

The \blockfinder for \rawblocks looks for a pair of 16-bit length and 16-bit bit-wise negated length contained in the header as illustrated in \cref{fig:deflate_blocks}.
The false positive rate for this check by itself is once every $2^{16}\SI{}{\byte} = \SI{64}{\kibi\byte}$ because the length is byte-aligned and can be of any length including $0$, which leaves the 16-bit bit-wise negated length to be checked.

The false positive rate is further reduced by also checking the block type, requiring the final-block bit to be 0, and requiring the padding to achieve byte-alignment to be filled with zeros.
Gzip compressed files with non-zero padding were not encountered.
Applying the \rawblock finder on 12 samples of \SI{1}{\gibi\byte} of random data, yields \SI{2040(90)}{} false positives on average, i.e., once every \SI{514(23)}{\kibi\byte} specified with one standard deviation.

In comparison to the \dynblock finder described in \cref{sct:dynblocks}, the false positive rate is higher.
But, this does not slow down overall performance because decompression of \rawblocks consists of a fast \texttt{memcpy} followed by a check of the next block header, which is likely to fail for a false positive.
Searching for \rawblocks is $7\times$ faster than the \blockfinder for \dynblocks, see \cref{fig:components-benchmarks} in \cref{sct:evaluation}.

Care has to be taken when matching the requested bit offset with the block found by the \rawblock finder.
Bit offsets of \rawblocks can be ambiguous because the length of the zero-padding is not known.
The preceding bits for the block type and the non-final bit are also zero and thus indistinguishable from the zero-padding.

\subsubsection{Compressed with Dynamic Huffman Codes}
\label{sct:dynblocks}

Finding \dynblocks is a bottleneck of the decoder because most \deflate blocks contain dynamic Huffman codes.
Checking these for correctness is more computationally intensive than for \rawblocks.
The steps for checking a \dynblock for correctness are in order:

\begin{enumerate}
    \item The final-block bit must be 0
    \item The two block-type bits must be ${01}_2$
    \item The five bits containing the number of literal/length codes must not be 30 or 31
    \item The \precode bit triplets must represent a valid and efficient \huffcode
    \item The data decoded with the \precode must not contain invalid backward pointers
    \item The Distance \huffcode must be valid and efficient
    \item The Literal \huffcode must be valid and efficient
\end{enumerate}

If any of these steps fail for a given offset, then that offset is filtered and the next offset will be tested.

\begin{figure}
    \centering
    \includegraphics[scale=1]{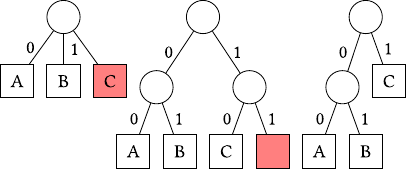}
    \caption{
        Three Huffman codes with invalid tree nodes highlighted in red.
        The code lengths for A, B, C are respectively given as (left) 1,1,1, (middle) 2,2,2, (right) 2,2,1.
        \textbf{Left:} The third symbol of bit length 1 is invalid because only two symbols can be encoded with a single bit.
        \textbf{Middle:} The code is inefficient because the Huffman code 11 is unused.
        \textbf{Right:} The code is valid. All available leaf nodes are used.
    }
    \label{fig:huffman-check}
\end{figure}

Huffman codes are stored as a list of code lengths per symbol.
Huffman codes will be invalid if there are more symbols with a given code length than the binary tree permits.
Huffman codes are not efficient if there are unused leaves in the binary tree.
\Cref{fig:huffman-check} illustrates the requirements on the Huffman code.

\begin{table}[htp]
    \centering
    \begin{tabular}{l|S[table-format=6.3(4)e2, separate-uncertainty]}
        Tested bit positions         & 1e12 \\
        \hline
        Invalid final block          & 500000.1(7)e6 \\
        Invalid compression type     & 375000.0(4)e6 \\
        Invalid \precode size         & 7812.47(14)e6 \\
        Invalid \precode code         & 77451.6(6)e6 \\
        Non-optimal \precode code     & 39256.9(4)e6 \\
        \hline
        Invalid \precode-encoded data & 386.66(5)e6 \\
        Invalid distance code        & 14.291(6)e6 \\
        Non-optimal distance code    & 77.126(16)e6 \\
        \hline
        Invalid literal code         & 340.6(10)e3 \\
        Non-optimal literal code     & 517.2(14)e3 \\
        \hline
        \hline
        Valid \deflate headers        & 202(27) \\
    \end{tabular}
    \caption{%
        Empirical filter frequencies listed top-down in the order they are checked.
        To get these frequencies, the \dynblock finder has been applied to \SI{1}{\tera\bit} + \SI{2300}{\bit} of data.
        It was chosen such that it can accommodate \SI{1e12} test positions that all have sufficient bits for the maximum possible \deflate header size.
        This simulation has been repeated \num{12} times.
        The uncertainties are specified with one standard deviation.
    }
    \label{tab:false_positives}
\end{table}

\Cref{tab:false_positives} shows empirical results for the amount of filtering for each of these checks.
It shows the importance of filtering as early as possible.

A lookup cache is used to speed up the first three checks.
Currently, this cache looks up a 14-bit value and returns an 8-bit number representing the offset of a potential \dynblock.
If 0 is returned, the \precode bits will be checked next.

Seeking inside the bit stream is computationally expensive if it requires refilling the internal buffer.
A 14-bit large buffer for the lookup table and a 61-bit large buffer for the \precode are maintained to reduce seeking.
The \precode check itself has also been optimized to filter as soon as the symbol length frequencies have been calculated.
It uses lookup tables and bit-level parallelism to speed up the computation of the frequency histograms.

Consider this example of adding two independent numbers, each packed into four bits, using only a single 8-bit addition instead of two separate 4-bit additions:
\begin{equation*}
    0101\,1101_2 + 0001\,0001_2 = 0110\,1110_2
\end{equation*}
The addition of the two 4-bit numbers in the high bits on the left is independent of the low bits if and only if the sum of the two 4-bit numbers in the lower bits does not overflow the range that can be represented with 4 bits.

The frequency histogram uses 5 bits per frequency, which is sufficient to avoid overflows because there are only up to 19 \precode symbols.
This is a total of 40 bits because there are 8 different code lengths from 0 to 7.
This bit packing also enables the usage of another lookup table for testing the histogram validity.
This table takes 20 consecutive bits corresponding to the frequencies for code lengths 1 to 5 as input.

This is followed by a short loop to filter all invalid or non-optimal \huffcodes, see \cref{fig:huffman-check}.
All lookup tables are computed at compile-time by making frequent use of the extended C++17 \texttt{constexpr} keyword support.
After this lookup table check is passed, the \huffcode structure for the \precode has to be created in a separate step.
This is partly duplicated work but it is only done in the unlikely case that the checks succeed.
We also found that only 1526 \precode frequency histograms belong to valid Huffman codes.
We precalculated this list of valid histograms in an attempt to reduce the histogram validity check to a simple lookup.
This did not improve performance.

The \huffcodes for the \deflate literals and distance codes are only initialized after both were found to be valid in order to speed up the preemptive filtering further.
This also leads to some duplicate work if both \huffcodes are valid, which, as shown in \cref{tab:false_positives}, only happens once in 5 billion.
See \cref{sct:evaluation} for a comparison of the effect of these optimizations.

\subsubsection{Compressed with Fixed Huffman Codes}

The implemented \blockfinder does not try to find \fixedblocks because only the final-block bit and the two block type bits can be checked without starting the decompression.
Further checks on the length or contents of the decompressed stream could be imposed but those would make assumptions that are not valid for all gzip-compressed files.
Fortunately, these types of blocks are rare and in practice only used for very small files or end-of-stream blocks that contain only a few bytes of data.
In the worst case of a gzip file only containing \fixedblocks, this would result in only one thread decompressing the file while the others threads try and fail to find valid blocks.

\subsubsection{Blocked GNU Zip Format}

The Blocked GNU Zip Format (BGZF)~\cite{htslib} splits the data into fixed-size streams and uses the gzip extra field~\cite{RFC1952} to store the decompressed stream size.
This makes it trivial to gather \deflateblock offsets to start parallel decompression from.
Furthermore, such start offsets do not require the decoded data from the previous blocks.
Because of this, the two-stage decoding can be skipped and parallel BGZF file decompression is trivial.
The \texttt{GzipChunkFetcher} contains specialized fast paths if a BGZF file has been detected.

\section{Evaluation}
\label{sct:evaluation}
In this section, we present single-core benchmarks for the components of our implementation followed by parallel decompression benchmarks.
All benchmarks in this section were executed on a single exclusively allocated \anon[cluster node]{AMD Rome cluster node provided by the Technische Universität Dresden}.
Each cluster node contains \SI{512}{\gibi\byte} of main memory and \num{2} AMD EPYC CPU 7702 processors with \num{64} cores each and with enabled simultaneous multithreading.
This yields a total of \num{256} virtual cores with a base clock frequency of \SI{2}{\giga\hertz}.

\subsection{Bit Reader}

\begin{figure}
    \centering
    \includegraphics[width=\linewidth]{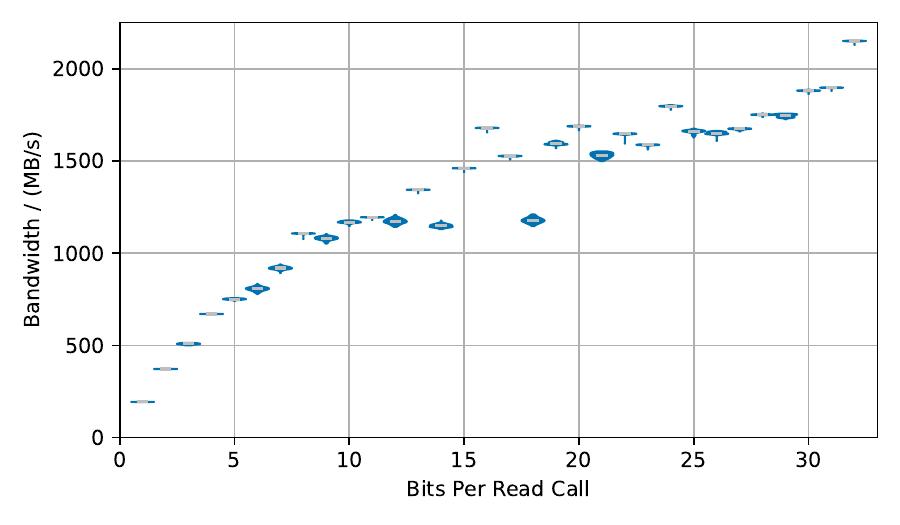}
    \caption{
        The \texttt{BitReader::read} method has been called with the given number of bits in a loop until the end of the data has been reached.
        The benchmark is single-threaded.
        The violin plot displays the sampled results from repeating the benchmarks 100 times.
        The data used for testing measures \SI{2}{\mebi\byte} scaled by the bits-per-read to give approximately equal benchmark runtimes.
    }
    \label{fig:bitreader}
\end{figure}

\Cref{fig:bitreader} shows that the performance of the \texttt{BitReader} class increases with the number of requested bits per call.
Thus, the bit reader should be queried as rarely as possible with as many bits as possible for optimal performance.
The number of requested bits depends on the Huffman decoder and the \blockfinder implementation.
One of the implemented Huffman decoders always requests the maximum \huffcode length, which is 15 bits for \deflate.

The parallelized gzip decompressor uses separate \texttt{BitReader} instances in each thread.
This increases the aggregated bit reader bandwidth accordingly.
Comparing the compressed bandwidth in \cref{fig:bitreader} and the decompressed bandwidth in \cref{fig:parallel-decompression} shows that the bit reader is not the main bottleneck even for the worst case with a compression ratio of 1.

\subsection{File Reader}

\begin{figure}
    \centering
    \includegraphics[width=\linewidth]{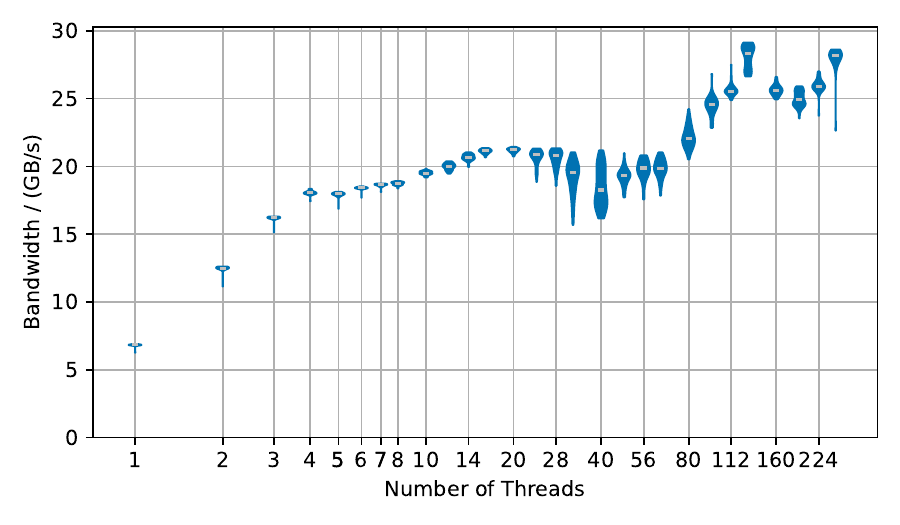}
    \caption{
        The \texttt{SharedFileReader} class is used from a set number of threads to read the file contents of a \SI{1}{\gibi\byte} large file located in \texttt{/dev/shm} in a strided manner.
        The implementation uses POSIX pread to read in parallel from the file.
        Each thread reads a \SI{128}{\kibi\byte} chunk, skips over the subsequent chunks read by other threads, and then reads the next chunk.
        The violin plot displays the sampled results from repeating the benchmarks 100 times.
    }
    \label{fig:filereader}
\end{figure}

\Cref{fig:filereader} shows results for multiple threads reading the same file residing in the in-memory file system \texttt{/dev/shm} in parallel.
The file was created by the main thread that has been pinned to core 0.
The n-th reader thread is pinned to the n-th core.
Each NUMA domain spans 16 physical cores.
The first NUMA domain consists of cores 0--15 and their simultaneous multi-threading (SMT) counterparts, cores 128--143.

\SI{18}{\giga\byte/\second} are reached reliably with 4 threads or more.
This bandwidth on the compressed input data is close to the highest decompression bandwidths shown in \cref{fig:parallel-decompression}.
It follows that file reading starts to become a bottleneck for more than 128 cores.
There is a slight decline in the file reading bandwidth for more than 20 threads, which is caused by the fixed work distribution.
The work distribution in \pragzip is dynamic to avoid such load balance issues.
The bandwidth increases with more than 64 threads because the additional threads reside on the second processor socket.

\subsection{Block Finder}

\begin{table}
    \centering
    \begin{tabular}{l|S[separate-uncertainty, table-align-uncertainty=true, table-format=4.4(4)]}
        Benchmark & {Bandwidth / (\SI{}{\mega\byte/\second})} \\
        \hline
        DBF zlib            & 0.1234 +- 0.0003 \\
        DBF custom deflate  & 3.403  +- 0.007  \\
        Pugz block finder   & 11.3   +- 0.7    \\
        DBF skip-LUT        & 18.26  +- 0.03   \\
        DBF rapidgzip       & 43.1   +- 1.1    \\
        \hline
        NBF                 & 301.8  +- 0.5    \\
        Marker replacement  & 1254   +- 6      \\
        Write to /dev/shm/  & 3799   +- 4      \\
        Count newlines      & 9550   +- 5      \\
    \end{tabular}
    \caption{
        The decompression bandwidths for different implementations of the \dynblock finder (DBF) and other components such as the \rawblock finder (NBF) and marker replacement.
        The benchmarks have been repeated 100 times.
        Uncertainties are given with one standard deviation.
        "DBF custom Deflate" is a trial-and-error method that uses \Pragzip's custom-written \deflate implementation instead of zlib's.
    }
    \label{fig:components-benchmarks}
\end{table}

\Cref{fig:components-benchmarks} displays benchmark results for all involved components.
Several implementations of the \dynblock finder are included in the comparison in order to show the impact of optimizations.
The \blockfinder using zlib for a trial-and-error approach is the slowest of all block finders.
A custom \deflate parser that returns early on errors is $28\times$ faster.
The lookup table implementation described in~\cref{sct:dynblocks} is $6\times$ faster than the custom \deflate parser.
The \blockfinder for \rawblocks is $7\times$ faster than the fastest \dynblock finder implementation. It is faster because it only needs to check the block type bits and compare the length with the one's complement length stored as a checksum.

The geometric mean of the \dynblock finder and the \rawblock finder bandwidths is \SI{38}{\mebi\byte/\second}.
This value can be used as a stand-in for the combined \blockfinder.
It is \SI{88}{\percent} as fast as the \dynblock finder and $\num{3.3}\times$ faster than the \blockfinder in \pugz.
A slow \blockfinder can be counter-balanced by increasing the chunk size proportionally.
This comes at the cost of increased memory usage.
Furthermore, the minimum file size required for full parallelization increases proportionally to the chunk size.

\subsection{Parallel Decompression of Base64-Encoded Random Data}

\begin{figure}
    \centering
    \includegraphics[width=\linewidth]{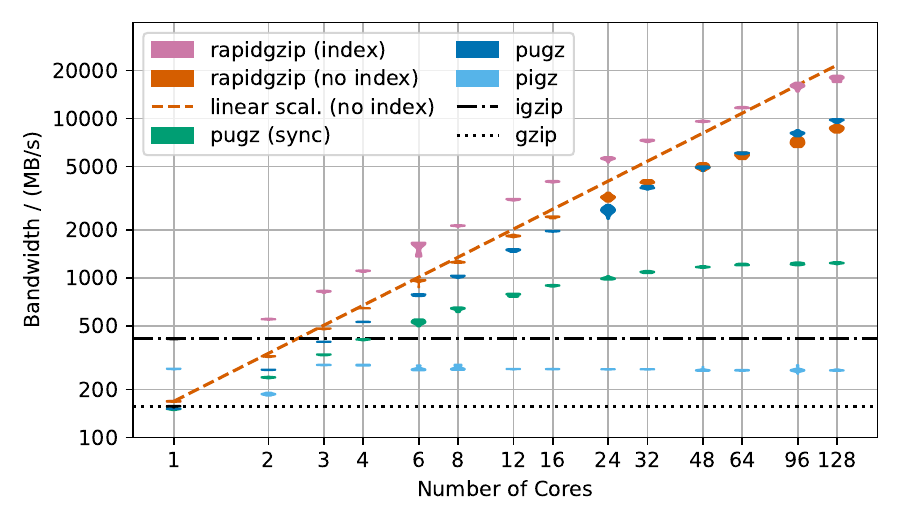}
    \caption{
        Benchmark results for gzip decompression.
        The results are written to \texttt{/dev/null} to avoid I/O write bottlenecks.
        It also makes the benchmarks comparable to those by \citeauthor{pugz}~\cite{pugz}.
        The degree of parallelism is enforced with the respective \texttt{-P} argument for \pragzip and \texttt{-t} for \pugz.
        Furthermore, the command line tool \texttt{taskset} is used to pin the process to a subset of the available cores.
        The file size for the base64 data was chosen to be \SI{128}{\mebi\byte} per core for \texttt{pugz (sync)},
          \SI{512}{\mebi\byte} per core for all other benchmarks of \pragzip and \pugz,
          and \SI{1}{\gibi\byte} for the \CLITOOL{gzip}, \CLITOOL{igzip}, and \pigz benchmarks.
        The violin plot displays the sampled results from repeating the benchmarks 20 times.
    }
    \label{fig:parallel-decompression}
\end{figure}

\cref{fig:parallel-decompression} shows the scaling behavior for parallel decompression for \pugz and \pragzip using base64-encoded random data compressed with \pigz.
\pigz was used to speed up benchmark setup times but triggered race conditions in \pugz.
Such crashes and deadlocks were reduced by setting the \pigz compression option \texttt{-{}-blocksize} to $\SI{4}{\mebi\byte}$ per degree of parallelization.
This option adjusts the workload assigned to each compression thread.
It does not adjust the \deflateblock size, which is \SI{75}{\kilo\byte} of compressed data on average.
Setting this option presumably reduces crashes because it reduces the number of empty \deflateblocks that are used to byte-align the \deflate streams that have been independently compressed by each thread.

\paragraph{Discussion of the Base64-Encoded Random Test Data}
The resulting file has a uniform data compression ratio of \num{1.315} and contains only a few backward pointers.
The compression ratio is mostly achieved with the Huffman coding.
The uniform compression ratio minimizes work distribution effects.
The low frequency of backward pointers makes it possible to fully resolve the markers in a \backrefwindow after a dozen kilobytes of data.
This enables the decoder to replace the two-stage method with single-stage decompression after a while.
The custom \deflate decoder is still used but further optimizations might delegate decompression to zlib in this case.
Therefore, this test data acts as a benchmark for all components except marker replacement.

\paragraph{\Pragzip Performance and Influence of the Index}
\Cref{fig:parallel-decompression} displays the benchmark results for \pragzip.
The benchmark is split into first-time decompression and decompression with an existing index file.
Decompression with an index file is generally faster because of these reasons:
\begin{itemize}
    \item marker replacement can be skipped
    \item the twice as large intermediate 16-bit format can be skipped
    \item decompression can be delegated to zlib
    \item the output buffer can be allocated beforehand in a single allocation because the decompressed size is known
    \item the workload is balanced because chunks in the index are chosen such that the decompressed chunk sizes are similar
\end{itemize}
For this case, \pragzip is the fastest when using more than one thread.
Using \num{128} cores, \pragzip reaches \SI{8.7}{\giga\byte/\second} without an index and \SI{17.8}{\giga\byte/\second} with an index.

\paragraph{Comparing \Pragzip with Pugz}

The bandwidth of \pragzip without a preexisting index is higher than \pugz for fewer than \num{64} cores.
Decompression with \pragzip is assumed to be slower for \num{64} cores and more because it returns the results in the correct order.
This behavior is the same as \texttt{pugz (sync)}, which is slower than \pragzip in all cases.
For this synchronized write mode, \pugz does not scale to more than 32 cores.
It achieves \SI{1.2}{\giga\byte/\second} decompression bandwidth for 48--128 cores.
For \num{128} cores, \pragzip without an index is $7\times$ faster than \texttt{pugz (sync)}.
The unsynchronized version of \pugz writes the decompressed data to the output file in undefined order.
This version has been benchmarked to reproduce the results in the work by \citeauthor{pugz}~\cite{pugz}.
The unsynchronized \pugz and \pragzip do scale to 128 cores but the parallel efficiency decreases.

\paragraph{Comparing \Pragzip with Single-Threaded Gzip Decompressors}

Using \num{128} cores, \pragzip achieves \SI{8.7}{\giga\byte/\second}, a speedup of \num{55} over decompression with gzip with \SI{157}{\mega\byte/\second}, and a speedup of \num{21} over \CLITOOL{igzip}, which decompresses with \SI{416}{\mega\byte/\second}.

When using a single thread, \pragzip reaches \SI{169}{\mega\byte/\second} and is slightly faster than \CLITOOL{gzip}.
It is twice as slow as \CLITOOL{igzip} and also slower than \pigz with \SI{270}{\mega\byte/\second}.
\CLITOOL{igzip} is the fastest single-threaded gzip decompressor.
The bandwidth of \pigz decreases for two cores and then increases again.
Pigz never outperforms \CLITOOL{igzip}.
It is not able to parallelize decompression but it offloads reading, writing, and checksum computation into separate threads.

\subsection{Parallel Decompression of the Silesia Corpus}

\begin{figure}
    \centering
    \includegraphics[width=\linewidth]{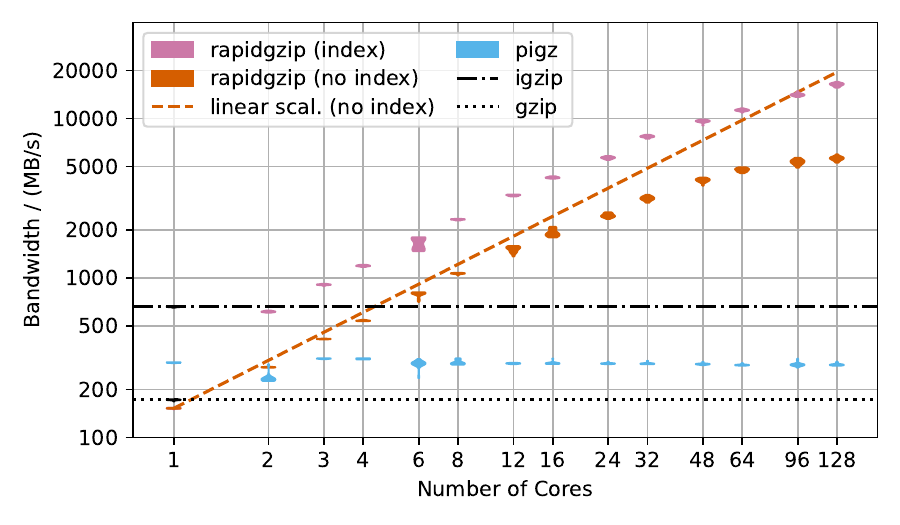}
    \caption{
        Benchmark results for gzip decompression of a compressed tarball of the Silesia dataset.
        The benchmark setup is the same as for \cref{fig:parallel-decompression}.
        The file size was scaled by concatenating 2 tarballs of Silesia per core.
        Compressing it with \pigz yields \SI{126}{\mega\byte} of compressed data per core.
        This is \SI{424}{\mega\byte} of uncompressed data per core with an average compression ratio of \num{3.1}.
    }
    \label{fig:parallel-decompression-silesia}
\end{figure}

The Silesia corpus~\cite{silesia} is commonly used to benchmark file compression.
\Cref{fig:parallel-decompression-silesia} contains benchmark results for parallel decompression using a gzip-compressed tarball of the Silesia corpus.
The Silesia tarball was concatenated repeatedly to itself to scale the file size with the degree of parallelism.
The comparison does not include \pugz because it is not able to decompress data containing bytes outside of the permitted range of 9--126.
It quits and returns an error when trying to do so.

\pragzip with an index reaches \SI{16.3}{\giga\byte/\second} and \SI{5.6}{\giga\byte/\second} without an index.
Compared to GNU gzip, which reaches \SI{172}{\giga\byte/\second}, this is a speedup of \num{95} and \num{33} respectively.
In comparison to \cref{fig:parallel-decompression}, it stops scaling after $\approx 64$ cores.
This is because the Silesia corpus contains more duplicate strings than base64-encoded random data.
The duplicate strings are compressed as backward pointers, which will be decompressed to markers in the first stage of decompression.
If there are no markers after \SI{32}{\kibi\byte}, the two-stage decompression can fall back to single-stage decompression.
This is the case for base64-encoded random data but not so for the Silesia corpus.
Between the two parallelized stages, the work distributor must resolve the markers in the last \SI{32}{\kibi\byte} of each chunk sequentially.
This limits efficient parallelization according to Amdahl's law.
Increasing the chunk size can alleviate this theoretically but will also increase the cache pressure.

The single-threaded decompression tools in \cref{fig:parallel-decompression-silesia} are faster than in \cref{fig:parallel-decompression} because the backward pointers generate decompressed data faster than undoing the expensive Huffman coding.
In the best case, such a backward pointer can copy 255 bytes from the window to generate a part of the decompressed stream.

\subsection{Parallel Decompression of FASTQ Files}

\begin{figure}
    \centering
    \includegraphics[width=\linewidth]{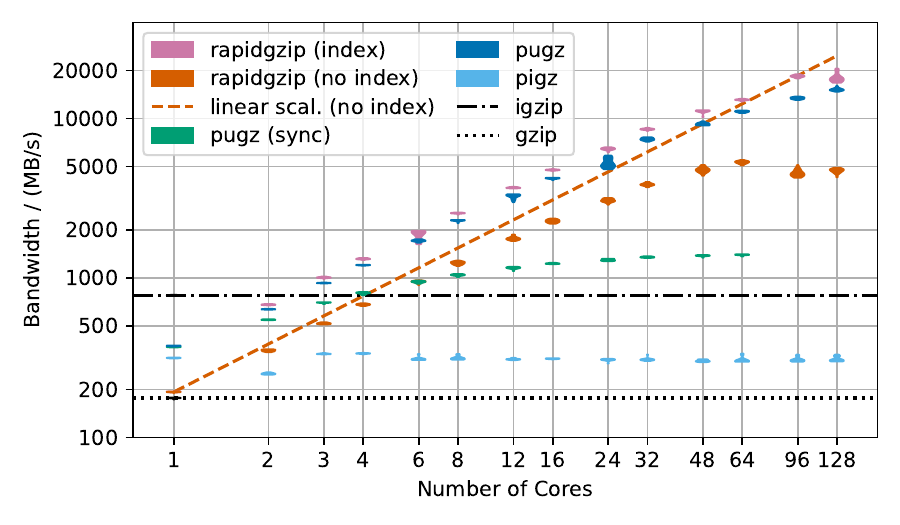}
    \caption{
        Benchmark results for gzip decompression of a compressed FASTQ file.
        The benchmark setup is the same as for \cref{fig:parallel-decompression}.
        The file size was scaled by concatenating two uncompressed FASTQ files per core.
        Compressing it with \pigz yields \SI{97}{\mega\byte} of compressed data per core.
        This is \SI{362}{\mega\byte} of uncompressed data per core with an average compression ratio of \num{3.74}.
    }
    \label{fig:parallel-decompression-fastq}
\end{figure}

\Cref{fig:parallel-decompression-fastq} contains benchmark results for parallel decompression on a FASTQ file\footnote{        \url{http://ftp.sra.ebi.ac.uk/vol1/fastq/SRR224/085/SRR22403185/SRR22403185_2.fastq.gz}}.
FASTQ files are used for storing biological sequences and metadata.
They were chosen for the benchmark because \pugz was developed to work with them.

Both, \pragzip with an existing index and \pugz without output synchronization scale up to 128 cores while \pragzip is slightly faster in all cases.
Without an index, \pragzip scales up to \num{48} cores and then stops scaling at \SI{4.9}{\giga\byte/\second} peak decompression bandwidth while \pugz with output synchronization scales up to 16 cores reaching a peak decompression bandwidth of \SI{1.4}{\giga\byte/\second}.
Benchmark results for \pugz with synchronization for \num{96} and \num{128} cores are missing because \pugz reproducibly threw errors when decompressing the test file with that many threads.
Overall, the scaling behavior shown in \cref{fig:parallel-decompression-fastq} is similar to that shown in \cref{fig:parallel-decompression-silesia} with the notable exception that \pragzip without an index stops scaling for more than \num{48} cores.

\subsection{Influence of the Chunk Size}

\begin{figure}
    \centering
    \includegraphics[width=\linewidth]{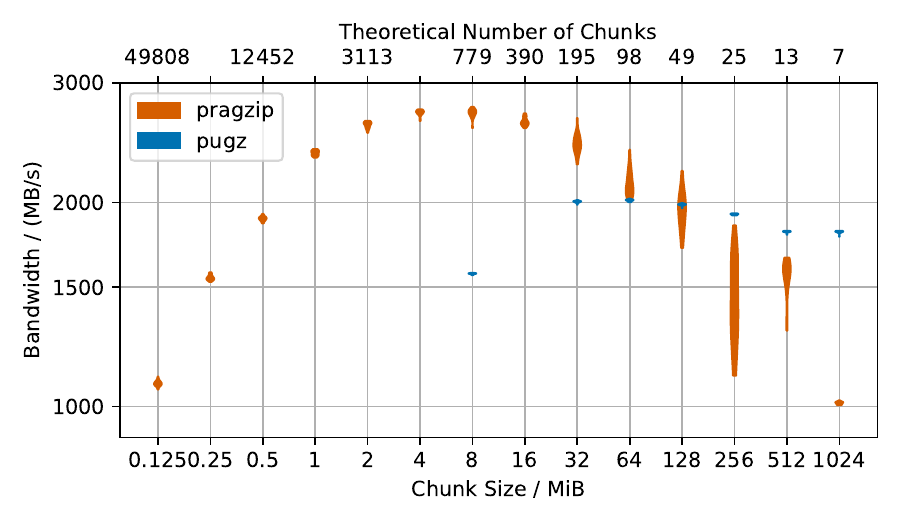}
    \caption{
        Benchmark results for gzip decompression unhindered by I/O write speeds using \num{16} cores and \SI{8}{\gibi\byte} of base64-encoded random data resulting in a gzip-compressed file size of \SI{6.08}{\gibi\byte}.
        The violin plot displays the sampled results from repeating the benchmarks 20 times.
    }
    \label{fig:chunk-size}
\end{figure}

The chunk size, i.e., the amount of compressed data in a work package submitted to the thread pool is one of the parameters that can be optimized.
\Cref{fig:chunk-size} shows that a very small chunk size leads to performance degradation because the overhead caused by the \blockfinder becomes too large.
A very large chunk size leads to performance loss because the work is not evenly distributed to the worker threads.
For \SI{512}{\mebi\byte} and more, the performance for \pugz stays stable because the maximum chunk size is limited to support even work distribution.
In these cases, each thread works on one chunk with a size of \SI{389}{\mebi\byte}.
This is not the case for \pragzip and performance degrades even more because not all worker threads have chunks to work on.
Assertion errors were observed with \pugz for chunks sized \SI{16}{\mebi\byte} and chunk sizes smaller than \SI{8}{\mebi\byte}.
The optimal chunk size for \pragzip is \SI{4}{\mebi\byte} and \SI{32}{\mebi\byte} for \pugz.
This is $8\times$ lower than \pugz owing to the optimized \blockfinder.
A lower chunk size reduces memory consumption.

\subsection{Influence of the Compressor}
\label{sct:gzip-compressors}

\begin{table}
    \centering
    \begin{tabular}{l|S[table-format=1.2]|S[separate-uncertainty, table-format=2.4(4)]}
        Compressor & {Compr. Ratio} & {Bandwidth / (\SI{}{\giga\byte/\second})}\\
        \hline & & \\[-1em]
        bgzip -l -1 & 2.99 &  5.65   +- 0.15   \\
        bgzip -l 0  & 1.00 & 10.6    +- 0.4    \\
        bgzip -l 3  & 2.81 &  5.90   +- 0.21   \\
        bgzip -l 6  & 2.99 &  5.67   +- 0.17   \\
        bgzip -l 9  & 3.01 &  5.64   +- 0.17   \\
        \hline & & \\[-1em]
        gzip -1     & 2.74 &  6.05   +- 0.15   \\
        gzip -3     & 2.90 &  5.55   +- 0.20   \\
        gzip -6     & 3.11 &  5.17   +- 0.15   \\
        gzip -9     & 3.13 &  5.03   +- 0.16   \\
        \hline & & \\[-1em]
        igzip -0    & 2.42 &  0.1586 +- 0.0013 \\
        igzip -1    & 2.71 &  6.15   +- 0.19   \\
        igzip -2    & 2.77 &  6.42   +- 0.14   \\
        igzip -3    & 2.82 &  6.52   +- 0.20   \\
        \hline & & \\[-1em]
        pigz -1     & 2.75 &  3.82   +- 0.07   \\
        pigz -3     & 2.91 &  3.81   +- 0.07   \\
        pigz -6     & 3.11 &  3.76   +- 0.09   \\
        pigz -9     & 3.13 &  3.73   +- 0.07   \\
    \end{tabular}
    \caption{
        The decompression bandwidths for decompressing Silesia with \pragzip using 128 cores.
        The benchmarks were repeated 20 times.
        Uncertainties are given with one standard deviation.
        The Silesia dataset was compressed with a variety of compression tools and compression levels as indicated in the first column.
        The Silesia dataset size was increased by concatenating 256 tarballs of the Silesia dataset.
        This creates a test file totaling \SI{54.2}{\giga\byte}.
    }
    \label{tab:compressors}
\end{table}

We have shown the weak-scaling behavior using \pigz with default compression levels to compress the data in the previous sections.
\Cref{tab:compressors} shows that \pragzip achieves efficient parallel decompression on the Silesia dataset for a wide variety of compression tools and compression levels.
This means that it can be used for almost any gzip-compressed data making it useful for cases in which the compression options cannot be adjusted.

\Cref{tab:compressors} also shows that there is significant variability across different compressors, e.g., decompression of \pigz-compressed files is slower than for \texttt{gzip}-compressed files.
One cause is the average \dynblock size, which can be chosen arbitrarily by the compressor.
There is only one Huffman coding per \dynblock.
The overhead required for reading and preprocessing that Huffman coding can be amortized better for longer \dynblocks.
Conversely, larger \dynblocks result in increased overhead for \pragzip to find the first \deflate block inside a chunk.

The dataset compressed with \texttt{bgzip -0} can be decompressed the fastest because, in this case, \texttt{bgzip} skips compression altogether and stores the data as \rawblocks.
Thus, \pragzip can fall back to a fast memory copy of the \deflate block contents.

The dataset created with \texttt{igzip -0} contains all the compressed data in a single \dynblock.
Therefore, the \pragzip decompression threads cannot find any other \deflate block boundary to start decompression from.
In this case, parallel decompression is not possible and the resulting decompression bandwidth is effectively a single-core decompression bandwidth.

\subsection{Comparison With Other Compression Formats}
\label{sct:compression-formats}

\begin{table}
    \centering
    \begin{tabular}{l|c|l|S[table-format=3]|S[separate-uncertainty, table-format=1.5(5), group-separator={}]}
        Com. & Rat. & Decompressor & {P} & {Bandw. / (\SI{}{\giga\byte/\second})}\\
        \hline & & & \\[-1em]
        bzip2 & 3.88 & lbzip2             & 1   &  0.04492 +- 0.00012 \\
        bgzip & 2.99 & bgzip              & 1   &  0.2977  +- 0.0023  \\
        gzip  & 3.11 & bgzip              & 1   &  0.2965  +- 0.0010  \\
        gzip  & 3.11 & rapidgzip          & 1   &  0.1527  +- 0.0010  \\
        gzip  & 3.11 & rapidgzip (index)  & 1   &  0.1528  +- 0.0007  \\
        gzip  & 3.11 & igzip              & 1   &  0.656   +- 0.009   \\
        zstd  & 3.18 & zstd               & 1   &  0.820   +- 0.006   \\
        zstd  & 3.18 & pzstd              & 1   &  0.816   +- 0.005   \\
        pzstd & 3.17 & pzstd              & 1   &  0.811   +- 0.003   \\
        lz4   & 2.10 & lz4                & 1   &  1.337   +- 0.013   \\
        \hline & & \\[-1em]
        bzip2 & 3.88 & lbzip2             & 16  &  0.667   +- 0.004   \\
        bgzip & 2.99 & bgzip              & 16  &  2.82    +- 0.07    \\
        gzip  & 3.11 & bgzip              & 16  &  0.3017  +- 0.0007  \\
        gzip  & 3.11 & rapidgzip          & 16  &  1.86    +- 0.12    \\
        gzip  & 3.11 & rapidgzip (index)  & 16  &  4.25    +- 0.03    \\
        pzstd & 3.17 & pzstd              & 16  &  6.78    +- 0.14    \\
        zstd  & 3.18 & pzstd              & 16  &  0.882   +- 0.006   \\
        \hline & & \\[-1em]
        bgzip & 2.99 & bgzip              & 128 &  5.5     +- 0.5     \\
        bzip2 & 3.88 & lbzip2             & 128 &  4.105   +- 0.024   \\
        gzip  & 3.11 & rapidgzip          & 128 &  5.13    +- 0.13    \\
        gzip  & 3.11 & rapidgzip (index)  & 128 & 16.43    +- 0.27    \\
        pzstd & 3.17 & pzstd              & 128 &  8.8     +- 0.8     \\
        pzstd & 3.17 & pzstd (no check)   & 128 &  8.8     +- 0.6     \\
    \end{tabular}
    \caption{
        The decompression bandwidths for a variety of decompression tools executed with fixed parallelization P.
        Uncertainties are given with one standard deviation.
        The first column shows the tool used for creating the compressed file when using default compression levels.
        The second column shows the compression ratio of the file.
        The file size was scaled by concatenating 2 tarballs of Silesia per core.
        This fixes the uncompressed file sizes to \SI{424}{\mega\byte}, \SI{3.39}{\giga\byte}, and \SI{27.13}{\giga\byte} for a parallelization of \num{1}, \num{16}, and \num{128} respectively.
    }
    \label{tab:compression-formats}
\end{table}

\begin{table}
    \centering
    \begin{tabular}{l|l||l|l}
        Software & Version & Software & Version \\
        \hline & & & \\[-1em]
        bgzip     & 1.17     &   pigz      & 2.7      \\
        gzip      & 1.12     &   pugz      & cc7c9942 \\
        igzip     & 9f2b68f0 &   pzstd     & 1.5.4    \\
        lbzip2    & 2.5      &   rapidgzip & d4aa8d4b \\
        lz4       & 1.9.4    &   zstd      & 1.5.4    \\
    \end{tabular}
    \caption{
        Software versions or git commit hashes used for benchmarks.
    }
    \label{tab:software-versions}
\end{table}

To conclude the benchmark section, we provide benchmark results for other compression formats in \cref{tab:compression-formats}.
Only default compression levels were included.
For a more comprehensive comparison of compression formats benchmarked with one core, we refer to lzbench~\cite{lzbench}.
Multiple things are of note in \Cref{tab:compression-formats}.

As can be seen for parallelization on 16 cores, both, \bgzip and \pzstd require specially prepared gzip and zstd files, respectively, to achieve parallelized decompression.
For \pzstd, Zstandard files with more than one frame are required but \CLITOOL{zstd} creates only Zstandard-compressed files with a single frame.
\Pragzip and \CLITOOL{lbzip2}~\cite{lbzip2} can work with arbitrary gzip and bzip2 files respectively.

Single-threaded \pragzip achieves \qty{153}{\mega\byte} decompression bandwidth while \bgzip is $\num{1.9}\times$ as fast and \CLITOOL{igzip} even $\num{4.3}\times$ as fast.
This shows further optimization potential for the custom-written \deflate implementation in \pragzip.

\CLITOOL{zstd} and \CLITOOL{pzstd} are $\num{5.4}\times$ faster than the single-core \pragzip but only $\num{1.2}\times$ faster than \CLITOOL{igzip} .
However, \CLITOOL{pzstd} cannot effectively use many cores.
The achieved speedup when using \num{16} cores instead of one core is \num{8.4}.
For \num{128} cores, the achieved speedup is \num{10.9}.
This ineffective parallelization closes the gap between \pragzip and \CLITOOL{pzstd} for larger core counts.
For 128 cores, \pragzip with an existing index becomes twice as fast as \CLITOOL{pzstd}.

The command line tool \CLITOOL{lbzip2} can be used for parallelized bzip2 decompression.
While \pzstd is $\num{18}\times$ faster than \CLITOOL{lbzip2} when using one core, it only is $\num{2.1}\times$ faster when using \num{128} cores.
Disabling checksum computation in \pzstd using the \texttt{-{}-no-check} option did not improve the decompression bandwidth.

The limited parallel scaling of \pzstd shows performance potential for the parallel architecture.
Alternatively, parallel Zstandard decompression could be implemented in the \pragzip architecture shown in \cref{fig:architecture} as has already been done for bzip2.

\section{Related Work}
\label{sct:related}

While various parallel tools for compression gzip in parallel exist, no tool for parallel decompression of arbitrary gzip files is known to us.

\paragraph{Parallel Gzip Compression}

The \texttt{bgzip} tool~\cite{htslib} divides the data to be compressed into chunks, compresses those in parallel as independent gzip streams, adds metadata, and concatenates the gzip streams into a Blocked GNU Zip Format (BGZF) file, which itself is a valid gzip file.
Alternatively, \pigz~\cite{pigz} compresses chunks in parallel as separate \deflate streams instead of gzip streams.
It uses workarounds such as empty \rawblocks to avoid having to bit-shift the results to fit the bit alignment of the previous \deflate stream when concatenating them.
There also exists a multitude of hardware implementations to speed up gzip compression for example for compressing network data~\cite{gzip_on_a_chip,qat,Ledwon2020HighThroughputFH}.

\paragraph{Modified Gzip File Formats Enabling Parallel Decompression}

Parallelization of gzip decompression is more difficult because, firstly, the position of each block is only known after decompressing the previous block.
Secondly, to decompress a \deflate block, the last \SI{32}{\kibi\byte} of decompressed data of the previous \deflateblock have to be known.

Some solutions like BGZF~\cite{htslib} work around these two issues by adjusting the compression to limit the dependency-introducing backward pointers, limiting Huffman code bit alignments, or storing additional Huffman code bit boundaries~\cite{gpu-decompresion}.
BGZF not only contains multiple independent gzip streams but those gzip streams also contain metadata storing the compressed size of each gzip stream to allow skipping over them.
Some hardware~\cite{cmos-gzip-decompression} and GPU~\cite{gpu-huffman} implementations speculatively decode Huffman codes ahead in the stream.
These implementations make use of the self-synchronizing nature of Huffman codes and reach decompressed data bandwidths of \SI{2.6}{\giga\byte/\second}~\cite{cmos-gzip-decompression}.

\paragraph{Two-Stage Decoding}

The closest work and the one that this work builds upon is \pugz~\cite{pugz}.
It uses a two-stage decompression scheme, which has been summarized in \cref{sct:theory:two-stage-decompression} and has been implemented in \pragzip.

\section{Conclusion}
\label{sct:discussion}

We have developed an architecture for parallel decompression and seeking in gzip files based on a cache-and-prefetch architecture.
This architecture is robust against false positives returned by the block finder.
This robustness enables us to extend the parallel gzip decompression algorithm presented by \citeauthor{pugz}~\cite{pugz}, which was limited to files containing bytes in the range 9--126, to all kinds of gzip-compressed files.
From this new implementation also emerge other improvements such as load balancing and support for multi-part gzip files.
The dynamic work distribution also improves upon the observed slowdowns of \pugz when writing the result to a file in the correct order.

We have implemented this architecture in the command line tool and library \pragzip.
We have achieved decompression bandwidths of \SI{8.7}{\giga\byte/\second} for base64-encoded random data,
  and \SI{5.6}{\giga\byte/\second} for the Silesia dataset when using 128 cores for parallelization.
This is $\num{21}\times$ and $\num{8.5}\times$ faster, respectively, than \CLITOOL{igzip}~\cite{igzip}, the fastest single-threaded gzip decompression tool known to us.
When compared to GNU gzip version 1.12, it is $\num{55}\times$ and $\num{33}\times$ faster, respectively.

In the future, we intend to add checksum computation and further optimizations to the custom \deflate implementation.
We will also address the limitations mentioned in~\cref{sct:limitations}.
In particular, the work splitting of chunks with very high decompression ratios will reduce the maximum memory requirements and further improve load balancing.

\begin{acks}
The authors are grateful to the Center for Information Services and High Performance Computing [Zentrum für Informationsdienste und Hochleistungsrechnen (ZIH)] at TU Dresden for providing its facilities for high throughput calculations.
\end{acks}

\bibliographystyle{ACM-Reference-Format}
\bibliography{bibliography}


\begin{thebibliography}{34}


\ifx \showCODEN    \undefined \def \showCODEN     #1{\unskip}     \fi
\ifx \showDOI      \undefined \def \showDOI       #1{#1}\fi
\ifx \showISBNx    \undefined \def \showISBNx     #1{\unskip}     \fi
\ifx \showISBNxiii \undefined \def \showISBNxiii  #1{\unskip}     \fi
\ifx \showISSN     \undefined \def \showISSN      #1{\unskip}     \fi
\ifx \showLCCN     \undefined \def \showLCCN      #1{\unskip}     \fi
\ifx \shownote     \undefined \def \shownote      #1{#1}          \fi
\ifx \showarticletitle \undefined \def \showarticletitle #1{#1}   \fi
\ifx \showURL      \undefined \def \showURL       {\relax}        \fi
\providecommand\bibfield[2]{#2}
\providecommand\bibinfo[2]{#2}
\providecommand\natexlab[1]{#1}
\providecommand\showeprint[2][]{arXiv:#2}

\bibitem[Abdelfattah et~al\mbox{.}(2014)]%
        {gzip_on_a_chip}
\bibfield{author}{\bibinfo{person}{Mohamed~S. Abdelfattah},
  \bibinfo{person}{Andrei Hagiescu}, {and} \bibinfo{person}{Deshanand Singh}.}
  \bibinfo{year}{2014}\natexlab{}.
\newblock \showarticletitle{Gzip on a Chip: High Performance Lossless Data
  Compression on {FPGA}s Using {OpenCL}}. In
  \bibinfo{booktitle}{\emph{Proceedings of the International Workshop on OpenCL
  2013 \& 2014}} (Bristol, United Kingdom) \emph{(\bibinfo{series}{IWOCL
  '14})}. \bibinfo{publisher}{Association for Computing Machinery},
  \bibinfo{address}{New York, NY, USA}, Article \bibinfo{articleno}{4},
  \bibinfo{numpages}{9}~pages.
\newblock
\showISBNx{9781450330077}
\urldef\tempurl%
\url{https://doi.org/10.1145/2664666.2664670}
\showDOI{\tempurl}


\bibitem[Adler(2022)]%
        {pigz}
\bibfield{author}{\bibinfo{person}{Mark Adler}.}
  \bibinfo{year}{2022}\natexlab{}.
\newblock \bibinfo{booktitle}{\emph{Pigz: A parallel implementation of gzip for
  modern multi-processor, multi-core machines}}.
\newblock
\urldef\tempurl%
\url{https://github.com/madler/pigz}
\showURL{%
\tempurl}


\bibitem[Biggers(2022)]%
        {libdeflate}
\bibfield{author}{\bibinfo{person}{Eric Biggers}.}
  \bibinfo{year}{2022}\natexlab{}.
\newblock \bibinfo{booktitle}{\emph{libdeflate: Heavily optimized library for
  {DEFLATE}/zlib/gzip compression and decompression}}.
\newblock
\urldef\tempurl%
\url{https://github.com/ebiggers/libdeflate}
\showURL{%
\tempurl}


\bibitem[Bonfield et~al\mbox{.}(2021)]%
        {htslib}
\bibfield{author}{\bibinfo{person}{James~K. Bonfield}, \bibinfo{person}{John
  Marshall}, \bibinfo{person}{Petr Danecek}, \bibinfo{person}{Heng Li},
  \bibinfo{person}{Valeriu Ohan}, \bibinfo{person}{Andrew Whitwham},
  \bibinfo{person}{Thomas Keane}, {and} \bibinfo{person}{Robert~M. Davies}.}
  \bibinfo{year}{2021}\natexlab{}.
\newblock \showarticletitle{{{HTSlib}: {C} library for reading/writing
  high-throughput sequencing data}}.
\newblock \bibinfo{journal}{\emph{GigaScience}} \bibinfo{volume}{10},
  \bibinfo{number}{2} (\bibinfo{date}{02} \bibinfo{year}{2021}).
\newblock
\showISSN{2047-217X}
\urldef\tempurl%
\url{https://doi.org/10.1093/gigascience/giab007}
\showDOI{\tempurl}
\showeprint{https://academic.oup.com/gigascience/article-pdf/10/2/giab007/36332285/giab007.pdf}
\newblock
\shownote{giab007}.


\bibitem[Boutell et~al\mbox{.}(1997)]%
        {png}
\bibfield{author}{\bibinfo{person}{Thomas Boutell} {et~al\mbox{.}}}
  \bibinfo{year}{1997}\natexlab{}.
\newblock \bibinfo{booktitle}{\emph{{PNG} (Portable Network Graphics)
  Specification Version 1.0}}.
\newblock \bibinfo{type}{RFC} 2083. \bibinfo{institution}{RFC Editor}.
\newblock
\urldef\tempurl%
\url{https://doi.org/10.17487/RFC2083}
\showDOI{\tempurl}


\bibitem[Byna et~al\mbox{.}(2008)]%
        {mpi-io-caching}
\bibfield{author}{\bibinfo{person}{Surendra Byna}, \bibinfo{person}{Yong Chen},
  \bibinfo{person}{Xian-He Sun}, \bibinfo{person}{Rajeev Thakur}, {and}
  \bibinfo{person}{William Gropp}.} \bibinfo{year}{2008}\natexlab{}.
\newblock \showarticletitle{Parallel I/O prefetching using MPI file caching and
  I/O signatures}. In \bibinfo{booktitle}{\emph{SC '08: Proceedings of the 2008
  ACM/IEEE Conference on Supercomputing}}. \bibinfo{pages}{1--12}.
\newblock
\urldef\tempurl%
\url{https://doi.org/10.1109/SC.2008.5213604}
\showDOI{\tempurl}


\bibitem[Collet and Kucherawy(2021)]%
        {zstandard}
\bibfield{author}{\bibinfo{person}{Yann Collet} {and} \bibinfo{person}{Murray
  Kucherawy}.} \bibinfo{year}{2021}\natexlab{}.
\newblock \bibinfo{booktitle}{\emph{Zstandard Compression and the
  '{application/zstd}' Media Type}}.
\newblock \bibinfo{type}{RFC} 8878. \bibinfo{institution}{RFC Editor}.
\newblock
\urldef\tempurl%
\url{https://doi.org/10.17487/RFC8878}
\showDOI{\tempurl}


\bibitem[Crawl(2023)]%
        {commoncrawl}
\bibfield{author}{\bibinfo{person}{Common Crawl}.}
  \bibinfo{year}{2023}\natexlab{}.
\newblock \bibinfo{booktitle}{\emph{Common Crawl}}.
\newblock
\urldef\tempurl%
\url{https://commoncrawl.org/}
\showURL{%
Retrieved 2023-01-15 from \tempurl}


\bibitem[Deng et~al\mbox{.}(2021)]%
        {imagenet21k}
\bibfield{author}{\bibinfo{person}{Jia Deng}, \bibinfo{person}{Wei Dong},
  \bibinfo{person}{Richard Socher}, \bibinfo{person}{Li-Jia Li},
  \bibinfo{person}{Kai Li}, {and} \bibinfo{person}{Li Fei-Fei}.}
  \bibinfo{year}{2021}\natexlab{}.
\newblock \bibinfo{title}{{ImageNet21K} (Winter 2021 Release)}.
\newblock
\newblock
\urldef\tempurl%
\url{https://www.image-net.org/}
\showURL{%
Retrieved 2023-01-15 from \tempurl}


\bibitem[Deorowicz(2003)]%
        {silesia}
\bibfield{author}{\bibinfo{person}{Sebastian Deorowicz}.}
  \bibinfo{year}{2003}\natexlab{}.
\newblock \bibinfo{booktitle}{\emph{The Silesia corpus}}.
\newblock
\urldef\tempurl%
\url{https://sun.aei.polsl.pl//~sdeor/index.php?page=silesia}
\showURL{%
Retrieved 2023-01-15 from \tempurl}


\bibitem[Deutsch(1996a)]%
        {RFC1951}
\bibfield{author}{\bibinfo{person}{Peter Deutsch}.}
  \bibinfo{year}{1996}\natexlab{a}.
\newblock \bibinfo{booktitle}{\emph{{DEFLATE} Compressed Data Format
  Specification Version 1.3}}.
\newblock \bibinfo{type}{RFC} 1951. \bibinfo{institution}{RFC Editor}.
\newblock
\urldef\tempurl%
\url{https://doi.org/10.17487/RFC1951}
\showDOI{\tempurl}


\bibitem[Deutsch(1996b)]%
        {RFC1952}
\bibfield{author}{\bibinfo{person}{Peter Deutsch}.}
  \bibinfo{year}{1996}\natexlab{b}.
\newblock \bibinfo{booktitle}{\emph{{GZIP} file format specification version
  4.3}}.
\newblock \bibinfo{type}{RFC} 1952. \bibinfo{institution}{RFC Editor}.
\newblock
\urldef\tempurl%
\url{https://doi.org/10.17487/RFC1952}
\showDOI{\tempurl}


\bibitem[Ersek and Izdebski(2023)]%
        {lbzip2}
\bibfield{author}{\bibinfo{person}{Laszlo Ersek} {and} \bibinfo{person}{Mikolaj
  Izdebski}.} \bibinfo{year}{2023}\natexlab{}.
\newblock \bibinfo{booktitle}{\emph{Parallel bzip2 utility}}.
\newblock
\urldef\tempurl%
\url{https://github.com/kjn/lbzip2}
\showURL{%
\tempurl}


\bibitem[for Standardization(2016)]%
        {xlsx}
\bibfield{author}{\bibinfo{person}{International~Organization for
  Standardization}.} \bibinfo{year}{2016}\natexlab{}.
\newblock \bibinfo{booktitle}{\emph{Information technology -- Document
  description and processing languages -- Office Open {XML} File Formats-- Part
  1: Fundamentals and Markup Language Reference}}.
\newblock \bibinfo{type}{Standard}. \bibinfo{institution}{International
  Organization for Standardization}, \bibinfo{address}{Geneva, CH}.
\newblock


\bibitem[Gailly and Adler(2004)]%
        {zlib}
\bibfield{author}{\bibinfo{person}{Jean-loup Gailly} {and}
  \bibinfo{person}{Mark Adler}.} \bibinfo{year}{2004}\natexlab{}.
\newblock \showarticletitle{Zlib compression library}.
\newblock  (\bibinfo{year}{2004}).
\newblock


\bibitem[Gill and Bathen(2007)]%
        {gill2007amp}
\bibfield{author}{\bibinfo{person}{Binny~S. Gill} {and} \bibinfo{person}{Luis
  Angel~D. Bathen}.} \bibinfo{year}{2007}\natexlab{}.
\newblock \showarticletitle{{AMP}: Adaptive Multi-stream Prefetching in a
  Shared Cache}. In \bibinfo{booktitle}{\emph{Proceedings of the 5th USENIX
  Conference on File and Storage Technologies (FAST’07)}},
  Vol.~\bibinfo{volume}{7}. \bibinfo{pages}{185--198}.
\newblock


\bibitem[Huffman(1952)]%
        {huffman}
\bibfield{author}{\bibinfo{person}{David~A. Huffman}.}
  \bibinfo{year}{1952}\natexlab{}.
\newblock \showarticletitle{A Method for the Construction of Minimum-Redundancy
  Codes}.
\newblock \bibinfo{journal}{\emph{Proceedings of the IRE}}
  \bibinfo{volume}{40}, \bibinfo{number}{9} (\bibinfo{year}{1952}),
  \bibinfo{pages}{1098--1101}.
\newblock
\urldef\tempurl%
\url{https://doi.org/10.1109/JRPROC.1952.273898}
\showDOI{\tempurl}


\bibitem[{Intel Corporation}(2022a)]%
        {igzip}
\bibfield{author}{\bibinfo{person}{{Intel Corporation}}.}
  \bibinfo{year}{2022}\natexlab{a}.
\newblock \bibinfo{booktitle}{\emph{{Intel(R)} Intelligent Storage Acceleration
  Library}}.
\newblock
\urldef\tempurl%
\url{https://github.com/intel/isa-l}
\showURL{%
\tempurl}


\bibitem[{Intel Corporation}(2022b)]%
        {qat}
\bibfield{author}{\bibinfo{person}{{Intel Corporation}}.}
  \bibinfo{year}{2022}\natexlab{b}.
\newblock \bibinfo{title}{{Intel(R) QAT}: Performance, Scale, and Efficiency}.
\newblock
\newblock
\urldef\tempurl%
\url{https://www.intel.com/content/www/us/en/architecture-and-technology/intel-quick-assist-technology-overview.html}
\showURL{%
\tempurl}


\bibitem[Katz et~al\mbox{.}(2022)]%
        {zip}
\bibfield{author}{\bibinfo{person}{Phillip Katz} {et~al\mbox{.}}}
  \bibinfo{year}{2022}\natexlab{}.
\newblock \bibinfo{title}{{APPNOTE.TXT} - {.ZIP} File Format Specification}.
\newblock
\newblock
\urldef\tempurl%
\url{https://pkware.cachefly.net/webdocs/APPNOTE/APPNOTE-6.3.10.TXT}
\showURL{%
\tempurl}


\bibitem[Kerbiriou and Chikhi(2019)]%
        {pugz}
\bibfield{author}{\bibinfo{person}{Maël Kerbiriou} {and}
  \bibinfo{person}{Rayan Chikhi}.} \bibinfo{year}{2019}\natexlab{}.
\newblock \showarticletitle{Parallel Decompression of Gzip-Compressed Files and
  Random Access to {DNA} Sequences}. In \bibinfo{booktitle}{\emph{2019 IEEE
  International Parallel and Distributed Processing Symposium Workshops
  (IPDPSW)}}. \bibinfo{pages}{209--217}.
\newblock
\urldef\tempurl%
\url{https://doi.org/10.1109/IPDPSW.2019.00042}
\showDOI{\tempurl}


\bibitem[Knespel(2022)]%
        {ratarmount}
\bibfield{author}{\bibinfo{person}{Maximilian Knespel}.}
  \bibinfo{year}{2022}\natexlab{}.
\newblock \bibinfo{booktitle}{\emph{Ratarmount: Random Access Tar Mount}}.
\newblock
\urldef\tempurl%
\url{https://github.com/mxmlnkn/ratarmount}
\showURL{%
\tempurl}


\bibitem[Ledwon et~al\mbox{.}(2020)]%
        {Ledwon2020HighThroughputFH}
\bibfield{author}{\bibinfo{person}{Morgan Ledwon}, \bibinfo{person}{Bruce~F.
  Cockburn}, {and} \bibinfo{person}{Jie Han}.} \bibinfo{year}{2020}\natexlab{}.
\newblock \showarticletitle{High-Throughput {FPGA}-Based Hardware Accelerators
  for Deflate Compression and Decompression Using High-Level Synthesis}.
\newblock \bibinfo{journal}{\emph{IEEE Access}}  \bibinfo{volume}{8}
  (\bibinfo{year}{2020}), \bibinfo{pages}{62207--62217}.
\newblock
\urldef\tempurl%
\url{https://doi.org/10.1109/ACCESS.2020.2984191}
\showDOI{\tempurl}


\bibitem[McCarthy(2022)]%
        {indexed_gzip}
\bibfield{author}{\bibinfo{person}{Paul McCarthy}.}
  \bibinfo{year}{2022}\natexlab{}.
\newblock \bibinfo{booktitle}{\emph{{indexed\_gzip}: Fast random access of gzip
  files in {Python}}}.
\newblock
\urldef\tempurl%
\url{https://github.com/pauldmccarthy/indexed_gzip}
\showURL{%
\tempurl}


\bibitem[Oracle(2022)]%
        {jar}
\bibfield{author}{\bibinfo{person}{Oracle}.} \bibinfo{year}{2022}\natexlab{}.
\newblock \bibinfo{title}{{JAR} File Specification}.
\newblock
\newblock
\urldef\tempurl%
\url{https://docs.oracle.com/en/java/javase/13/docs/specs/jar/jar.html}
\showURL{%
\tempurl}


\bibitem[Park et~al\mbox{.}(2008)]%
        {park2008data}
\bibfield{author}{\bibinfo{person}{Bora Park}, \bibinfo{person}{Antonio
  Savoldi}, \bibinfo{person}{Paolo Gubian}, \bibinfo{person}{Jungheum Park},
  \bibinfo{person}{Seok~Hee Lee}, {and} \bibinfo{person}{Sangjin Lee}.}
  \bibinfo{year}{2008}\natexlab{}.
\newblock \showarticletitle{Data extraction from damage compressed file for
  computer forensic purposes}.
\newblock \bibinfo{journal}{\emph{International Journal of Hybrid Information
  Technology}} \bibinfo{volume}{1}, \bibinfo{number}{4} (\bibinfo{year}{2008}),
  \bibinfo{pages}{89--102}.
\newblock


\bibitem[Satpathy et~al\mbox{.}(2019)]%
        {cmos-gzip-decompression}
\bibfield{author}{\bibinfo{person}{Sudhir Satpathy}, \bibinfo{person}{Vikram
  Suresh}, \bibinfo{person}{Raghavan Kumar}, \bibinfo{person}{Vinodh Gopal},
  \bibinfo{person}{James Guilford}, \bibinfo{person}{Mark Anders},
  \bibinfo{person}{Himanshu Kaul}, \bibinfo{person}{Amit Agarwal},
  \bibinfo{person}{Steven Hsu}, \bibinfo{person}{Ram Krishnamurthy},
  \bibinfo{person}{Vivek De}, {and} \bibinfo{person}{Sanu Mathew}.}
  \bibinfo{year}{2019}\natexlab{}.
\newblock \showarticletitle{A {1.4GHz} {20.5Gbps} {GZIP} decompression
  accelerator in 14nm {CMOS} featuring dual-path out-of-order speculative
  {Huffman} decoder and multi-write enabled register file array}. In
  \bibinfo{booktitle}{\emph{2019 Symposium on VLSI Circuits}}.
  \bibinfo{pages}{C238--C239}.
\newblock
\urldef\tempurl%
\url{https://doi.org/10.23919/VLSIC.2019.8777934}
\showDOI{\tempurl}


\bibitem[Sitaridi et~al\mbox{.}(2016)]%
        {gpu-decompresion}
\bibfield{author}{\bibinfo{person}{Evangelia Sitaridi}, \bibinfo{person}{Rene
  Mueller}, \bibinfo{person}{Tim Kaldewey}, \bibinfo{person}{Guy Lohman}, {and}
  \bibinfo{person}{Kenneth~A. Ross}.} \bibinfo{year}{2016}\natexlab{}.
\newblock \showarticletitle{Massively-Parallel Lossless Data Decompression}. In
  \bibinfo{booktitle}{\emph{2016 45th International Conference on Parallel
  Processing (ICPP)}}. \bibinfo{pages}{242--247}.
\newblock
\urldef\tempurl%
\url{https://doi.org/10.1109/ICPP.2016.35}
\showDOI{\tempurl}


\bibitem[Skibiński(2022)]%
        {lzbench}
\bibfield{author}{\bibinfo{person}{Przemysław Skibiński}.}
  \bibinfo{year}{2022}\natexlab{}.
\newblock \bibinfo{title}{{lzbench is an in-memory benchmark of open-source
  {LZ77/LZSS/LZMA} compressors}}.
\newblock
\newblock
\urldef\tempurl%
\url{https://github.com/inikep/lzbench}
\showURL{%
\tempurl}


\bibitem[Smith(1982)]%
        {smith1982cache}
\bibfield{author}{\bibinfo{person}{Alan~Jay Smith}.}
  \bibinfo{year}{1982}\natexlab{}.
\newblock \showarticletitle{Cache memories}.
\newblock \bibinfo{journal}{\emph{ACM Computing Surveys (CSUR)}}
  \bibinfo{volume}{14}, \bibinfo{number}{3} (\bibinfo{year}{1982}),
  \bibinfo{pages}{473--530}.
\newblock


\bibitem[Standard(2021)]%
        {odt}
\bibfield{author}{\bibinfo{person}{OASIS Standard}.}
  \bibinfo{year}{2021}\natexlab{}.
\newblock \bibinfo{title}{Open Document Format for Office Applications
  ({OpenDocument}) Version 1.3}.
\newblock
\newblock
\urldef\tempurl%
\url{https://www.oasis-open.org/standards/}
\showURL{%
\tempurl}


\bibitem[Storer and Szymanski(1982)]%
        {lzss}
\bibfield{author}{\bibinfo{person}{James~A. Storer} {and}
  \bibinfo{person}{Thomas~G. Szymanski}.} \bibinfo{year}{1982}\natexlab{}.
\newblock \showarticletitle{Data Compression via Textual Substitution}.
\newblock \bibinfo{journal}{\emph{J. ACM}} \bibinfo{volume}{29},
  \bibinfo{number}{4} (\bibinfo{date}{oct} \bibinfo{year}{1982}),
  \bibinfo{pages}{928–951}.
\newblock
\showISSN{0004-5411}
\urldef\tempurl%
\url{https://doi.org/10.1145/322344.322346}
\showDOI{\tempurl}


\bibitem[Wei\ss{}enberger and Schmidt(2018)]%
        {gpu-huffman}
\bibfield{author}{\bibinfo{person}{Andr\'{e} Wei\ss{}enberger} {and}
  \bibinfo{person}{Bertil Schmidt}.} \bibinfo{year}{2018}\natexlab{}.
\newblock \showarticletitle{Massively Parallel Huffman Decoding on {GPU}s}. In
  \bibinfo{booktitle}{\emph{Proceedings of the 47th International Conference on
  Parallel Processing}} (Eugene, OR, USA) \emph{(\bibinfo{series}{ICPP 2018})}.
  \bibinfo{publisher}{Association for Computing Machinery},
  \bibinfo{address}{New York, NY, USA}, Article \bibinfo{articleno}{27},
  \bibinfo{numpages}{10}~pages.
\newblock
\showISBNx{9781450365109}
\urldef\tempurl%
\url{https://doi.org/10.1145/3225058.3225076}
\showDOI{\tempurl}


\bibitem[Ziv and Lempel(1977)]%
        {lz77}
\bibfield{author}{\bibinfo{person}{Jacob Ziv} {and} \bibinfo{person}{Abraham
  Lempel}.} \bibinfo{year}{1977}\natexlab{}.
\newblock \showarticletitle{A universal algorithm for sequential data
  compression}.
\newblock \bibinfo{journal}{\emph{IEEE Transactions on Information Theory}}
  \bibinfo{volume}{23}, \bibinfo{number}{3} (\bibinfo{year}{1977}),
  \bibinfo{pages}{337--343}.
\newblock
\urldef\tempurl%
\url{https://doi.org/10.1109/TIT.1977.1055714}
\showDOI{\tempurl}


\end{thebibliography}

\end{document}